\documentclass[12pt,preprint]{aastex}
\usepackage{rotating}
\usepackage{subfigure}
\usepackage{natbib}
\usepackage{longtable}
\usepackage{threeparttable}

\shorttitle{New Identifications of GC X-ray Sources}
\shortauthors{DeWitt et al.}
 
\begin{document}

\title{Three New Galactic Center X-ray Sources Identified with Near-Infrared Spectroscopy}

\author{Curtis DeWitt\altaffilmark{1,2,3}, Reba M.
  Bandyopadhyay\altaffilmark{3}, Stephen S.
  Eikenberry\altaffilmark{3,4}, Kris Sellgren\altaffilmark{5},Robert Blum\altaffilmark{6}, Knut
  Olsen\altaffilmark{6}, Franz E. Bauer\altaffilmark{7,8} and Ata
  Sarajedini\altaffilmark{3}}

\email{curtis.n.dewitt@nasa.gov}
\altaffiltext{1}{Department of Physics, University of California,
  Davis, CA 95616, USA}
\altaffiltext{2}{Mail Stop 211-1, NASA Ames Research Center, Moffett Field, CA 94035, USA}
\altaffiltext{3}{Department of Astronomy, University of Florida, 211
  Bryant Space Center, P.O. Box 112055, Gainesville, FL 32611, USA}
\altaffiltext{4}{University of Florida Research Foundation Professor of Astronomy}
\altaffiltext{5}{Department of Astronomy, The Ohio State University, 140 West 18th Avenue, Columbus, OH 43210, USA}
\altaffiltext{6}{National Optical Astronomy Observatories, Tucson, AZ
  85719, USA}
\altaffiltext{7}{Pontificia Universidad Cat\'{o}lica de Chile, Departamento
de Astronom\'{\i}a y Astrof\'{\i}sica, Casilla 306, Santiago 22, Chile}
\altaffiltext{8}{Space Science Institute, 4750 Walnut Street, Suite 205,
Boulder, Colorado 80301, USA}

\begin{abstract}
We have conducted a near-infrared spectroscopic survey of 47 candidate
counterparts to X-ray sources discovered by the \textit{Chandra} X-ray
Observatory near the Galactic Center (GC). Though a significant number of
these astrometric matches are likely to be spurious, we sought out
spectral characteristics of active stars and interacting binaries,
such as hot, massive spectral types or emission lines in order to
corroborate the X-ray activity and certify the authenticity of the
match. We present three new spectroscopic identifications, including a
Be high mass X-ray binary (HMXB) or a $\gamma$ Cassiopeiae (Cas) system, a
symbiotic X-ray binary and an O-type star of unknown luminosity
class. The Be HMXB/$\gamma$ Cas system and the symbiotic X-ray binary are
the first of their classes to be spectroscopically identified in the GC region.
\end{abstract}

\keywords{Galaxy: center - infrared: stars - X-rays: stars}
\section{Introduction}

The \textit{Chandra} X-ray Observatory has uncovered nearly 10,000
X-ray point sources within the central $2^{\circ}\times0^{\circ}.8$ of
the Galaxy. The surface density of these sources and their heavy
extinction together imply that the majority reside near the Galactic
Center (GC), in an area that projects to $280\times110$ pc at the GC
distance of 8 kpc \citep{reid1993,muno2003}.

Most of these sources have faint, hard X-ray emission usually only associated with rare
categories of objects, such as high mass X-ray binaries (HMXBs), low
mass X-ray binaries (LMXBs), cataclysmic variables (CVs) with highly
magnetized white dwarfs, colliding-wind binaries with blue supergiants
or Wolf-Rayet stars, and in some rare types of symbiotic binaries
\citep{luna2007,muno2009}. Therefore these deep and well-localized
observations of Galactic X-ray
sources present an opportunity to find new examples of rare and
valuable objects within the Galaxy. In addition the identification of the nature of these
systems will constrain models of their binary formation channels and
their accretion physics, as well as provide a probe of the stellar
evolutionary history in the inner Galactic Bulge and around Sgr
A$^{*}$.

The principal difficulties with finding optical/IR counterparts to the
GC X-ray sources are the interstellar extinction and high stellar
density. On average, the GC lies
behind a column of $N_{H}=6\times10^{22}$ cm$^{-2}$, or $A_{V}=33.5$
mag \citep{predehl1994,baganoff2003}, precluding observations in the
ultraviolet/optical bands or with X-rays with $E < 2$ keV. 
This inevitably leads to observing in the near-infrared (NIR), where the
magnitude of dust extinction is significantly smaller than for optical
light ($A_{K_{s}}=\frac {A_{V}}{16}$; \citet{nishiyama2008}) and where
there are many identifying spectral features for stars and accretion-powered systems.

The high stellar density imposes its own difficulties. Astrometric
matches between X-ray sources and NIR sources are frequently chance
alignments instead of true physical counterparts. Our simulations of the astrometric
matching between X-ray point sources and NIR point sources toward the
central $17'\times 17'$ region around Sgr A$^{*}$ have shown that
$89\pm 3\%$ of apparent NIR astrometric matches to the hard X-ray
\textit{Chandra} sources are spurious \citep{dewitt2010}. Thus a
spectrum of a NIR candidate counterpart is required to establish an
authentic match and identify the counterpart. 

To date there have been more than 30 spectroscopic identifications of GC X-ray
sources from the Muno 2009 catalog \citep{mikles2006,hyodo2008,mauerhan2007,mauerhan2010}. All have been
either O supergiant or Wolf-Rayet stars with apparent magnitudes of
$K_{s}<12.2$ mag. O/WR stars with similarly hard X-ray emission have
been seen from colliding wind binaries (CWBs) or supergiant
HMXBs. However, as direct evidence for binary companions is lacking in
most cases, a single star emission mechanism cannot be ruled
out. Sixteen of these spectroscopically identified
counterparts are located in the central $17'\times17'$ GC region in
which we conducted our spectroscopic survey described in this paper. 

Our aim was to continue these efforts and simultaneously extend the
search to different classes of X-ray source whose infrared
counterparts are fainter than the limits of the previously published
studies. In this paper, we describe our spectroscopic survey targeting
47 NIR candidate counterparts to X-ray sources with $8.5 <K_{s}<14$ mag. We present the
spectra for three true counterparts, including a Be high
mass X-ray binary (HMXB) or $\gamma$ Cas system, a
symbiotic X-ray binary and an O-type star
with undetermined luminosity class. The Be type X-ray source and the
symbiotic X-ray binary are the first of their class to be
spectroscopically identified
in the GC. Our discoveries increase the number of spectroscopically
identified X-ray sources in the inner $17'\times17'$ by 20 $\%$.

\section{Sample Selection}

Our primary objective was to identify X-ray sources in the catalog of
\cite{muno2009} that reside near the Galactic Center distance. We used
our 2137 source NIR/X-ray matched catalog to select targets. This
catalog contains 1565 hard X-ray source matches and 572 soft X-ray
source matches to the 4268 X-ray sources detected by \citet{muno2009} in the inner $17'\times17'$ GC region. 

The division between soft and hard X-ray sources was
made to distinguish X-ray sources with low extinction that are within
4 kpc of the Earth and sources with high extinction, which are likely
to lie at 4 kpc and beyond. We use a threshold hardness ratio of
HR$_{0}<-0.175$ for defining soft sources, where HR$_{0}\equiv
\frac{h-s}{h+s}$, $s$ is the net soft photon counts with
$0.5<E<2.0$ keV and $h$ is the net hard photon counts with
$2.0<E<3.3$ keV \citep{muno2009}. Hard X-ray sources will also
include foreground sources with intrinsically hard X-ray spectra. To
focus on the source matches likely to be located near the Galactic
Center distance we chose to target sources with heavily reddened
colors of $(H-K_{s})>1.0$ mag that were matched to hard X-ray
sources. These matches frequently represent chance alignments of the
X-ray and NIR source coordinates due to the high NIR source density along
the line of sight, which makes targeting the true NIR counterparts difficult.

In \citet{dewitt2010}, we identified sets of near-infrared/X-ray properties that
maximized the number of probable true counterparts over the spurious
matches. One of our most promising findings was a set of 69 heavily reddened NIR matches to hard X-ray sources with a
$47\pm 9\%$ likelihood of being true counterparts. All of these
sources had X-ray positional errors of $\sigma_{X}\leq~1.0$\arcsec,
detections in $J$, $H$ and $K_{s}$ bands and relatively bright X-ray fluxes ($f_{X}\ge 0.0001$ counts
s$^{-1}$). Ten of the 16 previously known NIR/X-ray
counterparts in central $17'\times17'$ GC region are included in this set of 69 sources (e.g. \citealt{mikles2006, mauerhan2010}), which gives us confidence that
our matching simulations successfully identify the
properties of true counterparts.
Since we did not wish to include previously known targets in our
search, we limited our selection to the 59 targets within this
set for which there are no published identifications. Adjusting for
the known targets in this set, the remaining 59
targets have a 38$\pm$9 $\%$ chance of being
authentic ([47$\%~\times~69$ sources $-$ 10 known sources] $\div$ 59$=$0.38).

We observed 17 of these 59 ``high probability'' targets. In addition we observed 23 X-ray hard,
heavily reddened matches with properties that
excluded them from the ``high
probability'' set of 69, such as nominally faint X-ray
brightness or the lack of a $J$-band detection. These
targets were also chosen by using the \citet{dewitt2010} results to enhance
the probability for finding true counterparts, but their
properties and associated probabilities are too heterogeneous to report
here. Lastly, because of the
limitations for the placement of individual long slit and MOS mask
observations, we observed 7 NIR/X-ray counterparts 
with reddening or X-ray hardness ratios that should not correspond to GC
distances. In total, we observed 47 targets for our spectroscopic campaign to
find NIR counterparts to X-ray sources toward the GC.

\section{Spectroscopic Observations and Reduction}
We used the Large Binocular Telescope (LBT) LUCIFER1 spectrograph
\citep{seifert2003} in multi-object spectroscopy (MOS) mode, and the SOAR/OSIRIS spectrograph \citep{depoy1993} in
long slit mode to take spectra of our selected candidates. The dates
and weather conditions for the LUCIFER1 observations and OSIRIS
observations are listed in Table \ref{tab:observations}.  In this
table we include the official IAU name of each X-ray source (e.g. CXOUGC
J174533.5-285540) and the corresponding record number within the
\textit{Chandra} X-ray source catalog from \citet{muno2009}. For the
sake of brevity, we use the X-ray record number, designated by the
prefix, ``XID'' for referring to X-ray sources and their NIR
counterparts.

\begin{table}
\tiny
  \caption{Spectroscopic Observations of the Newly Identified
    NIR Counterparts
    to GC \textit{Chandra} Sources.}
  \label{tab:observations} {\tiny\begin{tabular}{ r r l r l r r r}
      \hline
      \textit{Chandra} X-ray & X-ray Catalog & Date of      & Telescope and & R ($K$ band),                           & Airmass & Seeing & Sky        \\
       Source (CXOUGC) & Number       & Observations & Instrument    & $\frac{\lambda}{\delta\lambda}$ &         &        & Conditions \\
      \hline
      J174552.9-285537 & XID 3275 & 5/17/2010 & LBT/LUCIFER & 2500 & 2.06 & $1$\arcsec$.5$ & thin cirrus \\
      J174528.7-290942 & XID 6592 & 8/20/2010 & SOAR/OSIRIS & 1200 & 1.01 & $0$\arcsec$.5$ & clear \\
      J174528.8-285726 & XID  947 & 8/19/2010 & SOAR/OSIRIS  & 1200 & 2.09 & $0$\arcsec$.5$ & clear \\
      \hline
    \end{tabular}}
\end{table}

\subsection{LUCIFER1 Observations}

LUCIFER1 is a near-infrared spectrograph on the Large Binocular
Telescope at Mount Graham Observatory, Arizona that performs
multi-object spectroscopic (MOS) observations by use of laser-cut slit
masks. The field of view of each slit mask is
4\arcmin$\times$2\arcmin.5. For our observations, we selected the $HK$
grism which simultaneously provides $R=2500$ in the $K$ band and
$R=1900$ in the $H$ band. In order to obtain the full spectral range
targets must coincide with the center of the mask, since the $HK$
spectrum spans the entire 2048 pixel detector width. We selected
targets that would be positioned for coverage of the $\lambda=$2.1661 
$\mu$m Brackett $\gamma$ feature, so that this important accretion
signature would always be detectable if present.

We observed two MOS masks with LUCIFER on 17 May 2010. The MOS mask
positions were chosen to get the greatest possible number of the 59
``high-probability'' NIR/X-ray matches found in \citet{dewitt2010}. In
total,  we were able to position 9 of these sources on the two masks. The mask
center positions were $\alpha=$17:45:50, $\delta=$-28:57:29 (J2000) for
Mask 1 and $\alpha=$17:46:00, $\delta=$-28:54:24 (J2000) for Mask 2.

LUCIFER1 had only recently been commissioned for MOS mode at the time
of our observations. The procedures for mask alignment and guiding
were still being optimized and were expected to take up large amounts
of overhead time. Therefore, instead of using known A0V telluric
standards that would require recentering the slit mask, we used
unreddened sources in the MOS mask field of view which had NIR colors
consistent with stars earlier than G0. This approach has the
disadvantage that the exact spectral type of the reference star is
unknown in advance and must be derived as part of our spectroscopic
analysis , which is affected by the atmospheric transmission spectrum
that we desire to remove. We describe the procedure for correcting for
atmospheric transmission for LUCIFER1 in \S 3.2.2.

Mask 1 and Mask 2 were designed for 18 and 21 targets,
respectively. These targets included 19 of our NIR/X-ray candidate
counterparts, 3 telluric reference stars and also GC red giant
branch stars in support of another science project.
 
Each target slit was $0$\arcsec$.5 \times 6$\arcsec$.0$. During
observations, the MOS masks were aligned to bright reference stars
within the mask field of view and an ABBA nodding sequence containing
eight 100 second exposures was taken for each mask. The nod length was
2\arcsec.5. We attempted to select only targets that were free of
stars in the sky positions of the nod pattern in order to facilitate
clean sky subtraction. This was not possible for $\sim \frac{1}{2}$ of
the targets, and in these cases we selected targets with clear sky in
just one of the nod positions. As we discuss in \S 3.2.1, we discarded
the nod positions with contaminated skies.  The NIR/X-ray candidate counterpart targets on each mask
ranged from $10.5 <K_{s}<13.9$ mag. At the end of the night arc lamps
and spectral flat fields were taken with each MOS mask in place. Dark
exposures were acquired for all exposure times used in our
observations.

\subsection{LUCIFER1 data reduction}

We performed the LUCIFER1 data reduction using the Florida Analysis
Tool Born of Yearning for high quality data (FATBOY) software
package \citep{warner2008}.  We began the reduction process by
subtracting dark frames from all images to remove the detector dark
current and then dividing by spectral flat fields taken with each slit
mask to correct for the grating blaze function and the detector's
wavelength sensitivity. 

Then we subtracted the 2-D spectral frames of the B dither position from
the A dither position, pairing the A and B frames by the time they
were taken. The resultant ``sky-subtracted'' 2-D spectra should have
removed the bulk of the variable OH line emission, and largely contain
only signal from the target object. We
coadded all the A sky-subtracted spectra, coadded all the B
sky-subtracted spectra, and finally combined the A and B spectra by
multiplying the B component by -1 and shifting it in the slit
direction by the length of the dither. For some targets there was a contaminating star in the target slit
that was acquired at the sky position of one of the AB dither
positions. In these cases we discarded the contaminated positions from
the stacked spectrum. 

At this point we corrected the
curvature of the spectral line and continuum shapes of the 2-D spectra
by performing a rectification so that the spatial and spectral
channels of the 2-D were all aligned in x and y pixel space. We
extracted 1-D spectra by summing the 2-D spectra in the slit
direction.

At the sky subtraction phase, a median image was also created which
isolates the OH sky lines rather than removing them. We rectified this
sky line image using the same transformation as for the sky-subtracted
spectra. We calibrated the wavelength solution for each target
spectrum using a 2$^{nd}$ order polynomial fit to the centers of
atmospheric OH emission lines with known wavelength in the median
image. The errors on the wavelength calibration for each target were
between 0.03 and 0.1 \AA.

\subsubsection{Removal of Residual OH sky lines}

We encountered residual OH emission for some sources
that was not completely removed after sky subtraction. This was
most apparent in the faintest targets and when we were forced to use
only one object/sky nod position.  To treat these cases, we adapted
the methods of \citet{davies2007}. We extracted the center parts of
the median-stacked 2-D spectra (which isolated the OH sky spectra) to
generate a template of the OH sky-lines for each target spectrum. We
fit a power law to the thermal continuum in the sky template spectrum
and removed it. Then the sky lines were broken into groups which share
the same intensity variations \citep{davies2007}. These different
components of the sky template were then added or subtracted to
visually remove residual OH emission lines.

\subsubsection{Telluric Transmission Correction}
To produce telluric transmission spectra we needed to remove intrinsic
absorption lines from the telluric spectrum, which requires knowledge
of the spectral type. We chose targets with $(J-K_{s})< 0.1$ mag and
which were unblended in our \textit{ISPI} imaging. At this threshold
our telluric reference spectral types should have been limited to G0
types and earlier \citep{ducati2001}. O/B stars have He I absorption
features at $\lambda=2.112/2.113~\mu$m and $1.700~\mu$m. Neither of our
selected standards showed these features, which constrains the
spectral types to A0-G0. For these spectral types, the only strong
absorption features should be the Brackett series. Br $\gamma$ was
easily detected in $K$ band, as was Br 10-4 through Br 16-4 in $H$
band. We used a synthetic telluric absorption spectrum created for
Mauna Kea conditions at $sec(z)=2.0$ with \textit{ATRAN}
\citep{lord1992}. We adjusted this synthetic spectrum by adjusting its
depth and wavelength offset so that it would best remove
the deep telluric absorption bands at 2 $\mu$m and 2.05 $\mu$m in the $K$
band and 1.6 $\mu$m and 1.63 $\mu$m in the $H$ when it was divided from
our observed telluric reference spectra. This allowed us to
see, for the most part, the profiles of Brackett absorption features,
although the telluric line-corrected regions were still quite noisy due
to the usage of an artificially derived transmission spectrum.

We fit Br $\gamma$, and Br 10-4 through Br 16-4 using Gaussian
functions and removed them. Brackett $\delta$ was too contaminated by
atmospheric absorption to fit reliably. We used the ratio of depths
($1:1.2$) of Br $\gamma$/$\delta$ found in the synthetic spectrum of
the A0 V star Vega \citep{castelli1994} to remove the Brackett
$\delta$ line which is likely to exist in this region of our telluric
standard stars, but we consider the resultant signal from
$1.8<\lambda<1.95~\mu$m to be highly suspect. After these intrinsic
stellar absorption lines were removed, we restored the telluric lines
(which have been only partially removed by the synthetic telluric
spectrum), and normalized the resultant spectrum. At this point we had
a normalized telluric spectrum with the underlying stellar features
from which it was derived removed. We divided all our LUCIFER1 targets by this
telluric spectrum.

\subsection{OSIRIS observations}

We used OSIRIS on the 4-m SOAR telescope on Cerro Pachon, Chile in low
resolution $K$ band mode ($R=1200$) with the $1$\arcsec$\times 110$\arcsec$~$long
slit. The long slit length allowed us to target two candidates at once
by adjusting the slit position angle to acquire both star
positions. As often as possible, we tried to pair sources by matching
them closely in magnitude so that acceptable signal could be
reached simultaneously.  

In total, we observed 28 candidate counterparts to X-ray sources with
OSIRIS on August 19-20, 2010. We primarily targeted candidate
counterparts with $K_{s}<12.5$ mag. Eight targets were observed from
the list of 59 ``high probability'' targets that had not been
previously observed. The exposures were taken with ABBA
nodding, where the nod length was carefully chosen to avoid
contaminating stars in the sky positions for both targets on the slit.

We took a spectrum of the A1 V star HD160461 at least once per hour
for the purpose of telluric calibration. This star was never more than
0.05 different in airmass from any of our Galactic Center targets.

\subsection{OSIRIS data reduction}

The basic steps of the OSIRIS data reduction were handled with FATBOY
in the same way as for LUCIFER1, as described in \S 3.2.  We
calibrated the wavelength solution using both HeNeAr lamp exposures
from the beginning and end of the nights and the positions of OH
sky-lines acquired with target spectra. At the $R=1200$ resolution in
the $K$ band, the detectable OH lines do not span the full $K$ band
range. We calculated the wavelength solution with the arc lamp
exposures, and shifted the solution to the zero-positions of a set of
four of the brightest OH lines in the target spectra. The HeNeAr
wavelength solution was determined to have less than $0.2~$\AA$~$scatter
over both nights by measuring the difference in wavelength solutions
of arc lamps taken during the night.

\begin{figure}[ht]
\includegraphics[scale=0.75] {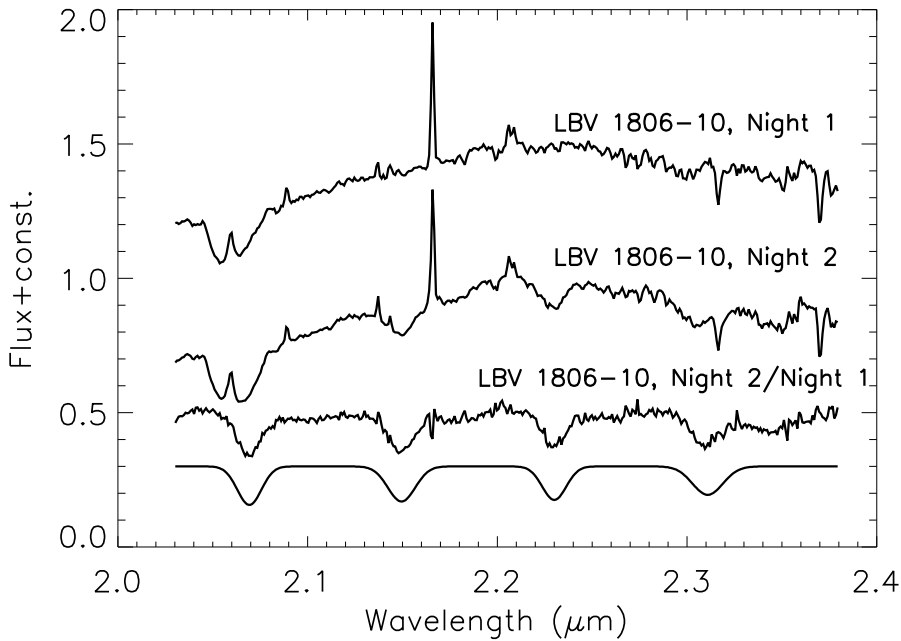}
\caption{OSIRIS $K$-band spectra of LBV 1806-20 taken on successive nights. In the
  second night the spectrum is afflicted by a regularly spaced pitting
  in the continuum. The lowest line shows our Gaussian fits to the
  ratio of the two LBV spectra. We used this pattern to correct
  pitting issues in the rest of our target spectra.}
  \label{fig:osirispitting}
\end{figure}

Many of our spectra were affected by the intermittent appearance of
regularly spaced dips in the spectrum. In Figure
\ref{fig:osirispitting} we show our spectra from LBV 1806-20 taken on
successive nights. We observed this source during the $\sim$20 minute
interval on both observing nights when the GC targets were beyond the azimuth limits of the SOAR
telescope. Our spectra of LBV 1806-20 closely resembles previously
published $K$-band spectra from this source, such as in \citet{eikenberry2004}. 
However, the second night spectrum shows four large
semi-regularly spaced dips in the continuum that were not present in the
first night spectrum. We observed similar features in roughly 50 $\%$
of our other OSIRIS targets. By visual examination we noted that the features
changed in amplitude and central position but the spacing and relative
amplitudes of the dips appeared to stay constant. 

This is a known but unsolved issue with OSIRIS since its installation
on SOAR from its previous location at the 4 m CTIO telescope
(R.Blum, \textit{priv. comm.}). We created a template for the shape of
these features by fitting Gaussian functions to the four pits in the
ratio of the night 2 and night 1 OSIRIS spectra of LBV 1806-10. The
pitting template is shown at the bottom of Figure
\ref{fig:osirispitting}. For each target that was affected we visually
matched the amplitude and position of the pitting template and divided
it from the target spectrum. We do not believe that our results are
significantly affected since the pitting features are broad compared
to the stellar absorption lines and emission lines. 

The telluric correction for OSIRIS was performed by first removing the
intrinsic Brackett $\gamma$ absorption from each of the telluric
standard spectra by
fitting a Gaussian function to the Brackett $\gamma$ position and
dividing it out. We then normalized the telluric transmission spectra
to 1.0. Each science target spectrum was divided by the normalized telluric
standard spectrum taken closest in time and airmass to the
science target observations.

\section{Results}
We identified three counterparts to hard X-ray
sources near the Galactic Center. As described in detail below, these
discoveries include a Be HMXB or $\gamma$ Cassiopeiae (Cas) system and symbiotic
X-ray binary. Both are the first of their types to be
spectroscopically identified within the $2^{\circ}\times0^{\circ}.8$ region toward Sgr A$^{*}$.
 The third newly discovered X-ray system is an O type star near the GC distance, which may be a
low luminosity HMXB, a CWB or an unusually X-ray hard single star. 

Within our 7 foreground targets, we uncovered an O type supergiant
counterpart to CXOUGC J174537.9-290134, confirming the spectroscopic
identification published by \citet{mauerhan2010b} shortly after our observations were taken.
Table \ref{tab:obssummary} summarizes our NIR spectroscopic identifications
of GC counterparts to X-ray sources. The other 6 foreground targets
were primarily found to be dwarf stars without emission lines. 

The 37 remaining GC target spectra were found to be late-type giants without emission lines. In these cases the
matches are either spurious or they represent true counterparts that
were targeted during a low-activity state. For reference, we include the coordinates and NIR photometry of all our 
spectroscopic targets in the Appendix.

\begin{table}[!ht]
\caption{Spectroscopically Identified Counterparts to X-ray Sources toward the GC.}
\label{tab:obssummary}
\begin{tabular}{ l l l cr}
\hline
  X-ray        & Spectral &X-ray   &Approximate \\
  Source Number & Type             & Source Type                             &Distance     \\ 
\hline
  3275 &  B0-3e III   & Be HMXB or $\gamma$ Cas system & GC distance \\
  6592 & M7 III         &  symbiotic XRB  & GC distance  \\
    947 & O I, III or V &  CWB, HMXB, single O star   &  GC distance\\
\hline
\end{tabular}
\end{table}

The ISPI magnitudes and colors of these three NIR counterparts and
their X-ray characteristics are presented in Table
\ref{tab:irphotometry}. We detail the
properties of each source below.

\begin{table}[!ht]
\begin{threeparttable}
\tiny
\setlength{\tabcolsep}{0.05in}
\caption{ISPI Near Infrared photometry and X-ray properties of newly
  identified systems in the GC \citep{dewitt2010}.}
\label{tab:irphotometry}
{\begin{tabular}{ l l l l l l l l l l l l l cr}
\hline
  X-ray                 & RA$_{IR}$   & DEC$_{IR}$ & $J$        & $H$      & $K_{s}$& Sep.\tnote{a} & $\sigma_{X}$\tnote{b} & HR$_{0}$  & Type & log(F$_{X}$) \\
  Source No.  & (J2000.0)  & (J2000.0)   & (mag)& (mag)& (mag)  & (\arcsec)             & (\arcsec)&              &          &                  \\
\hline
  3275 & 17:45:52.97    & -28:55:37.0 & 17.69$\pm$ 0.18  &  14.75$\pm$ 0.04   &  13.08$\pm$ 0.04 & 0.15 & 0.6 & $-9.0$\tnote{c}& hard & -3.968\\
  6592 & 17:45:28.79    & -29:09:42.8 &  -                           &  13.03$\pm$ 0.03   &9.88$\pm$ 0.03     & 0.2  & 0.6 & $0.716^{+0.285}_{-0.371}$  & hard & -2.673\\
   947 &  17:45:28.88    & -28:57:26.4 & 16.23$\pm$ 0.06  &  13.29$\pm$  0.03  &  11.61$\pm$ 0.03 & 0.06 & 0.5 & $1.0^{+0}_{-0.822}$    & hard & -4.110 \\
\hline
\end{tabular}}
\begin{tablenotes}
\item [a] Separation in arcseconds between the IR counterparts and the X-ray positions from \citet{muno2009}.
\item [b] X-ray 95$\%$ positional uncertainty from \citet{muno2009}.
\item [c] This value means that there were no significant counts in the 0.5-2 keV and 2-3.3 keV bands.
\end{tablenotes}
\end{threeparttable}
\end{table}

\subsection{XID 3275: a Galactic Center Be HMXB or $\gamma$ Cassiopeiae System}

\begin{figure}[ht]
\centering
\subfigure[] 
{
    \label{fig:subim_xid3275_J}
    \includegraphics[width=3.5cm]{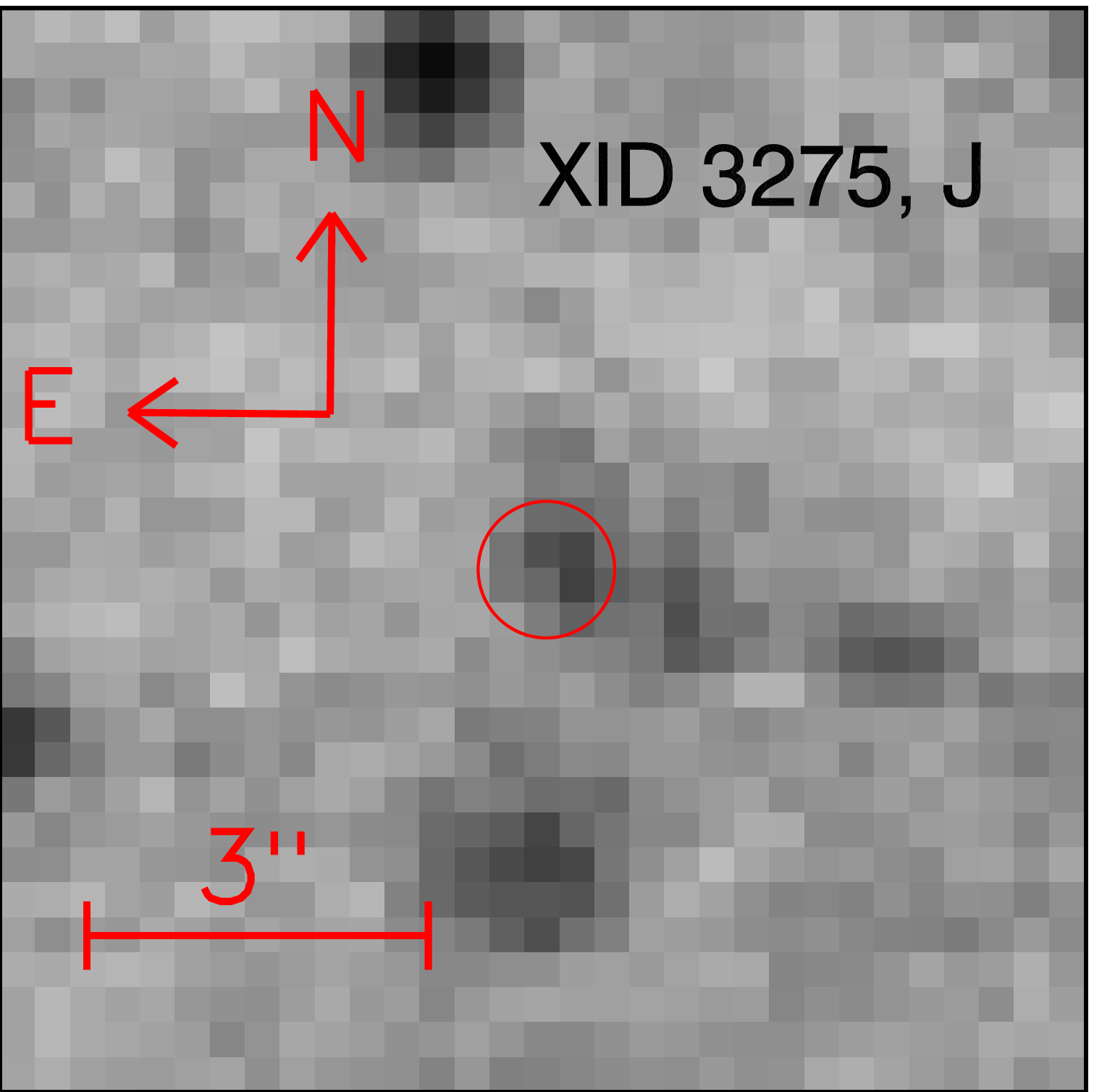}
}
\hspace{0.1cm}
\subfigure[] 
{
    \label{fig:subim_xid3275_H}
    \includegraphics[width=3.5cm]{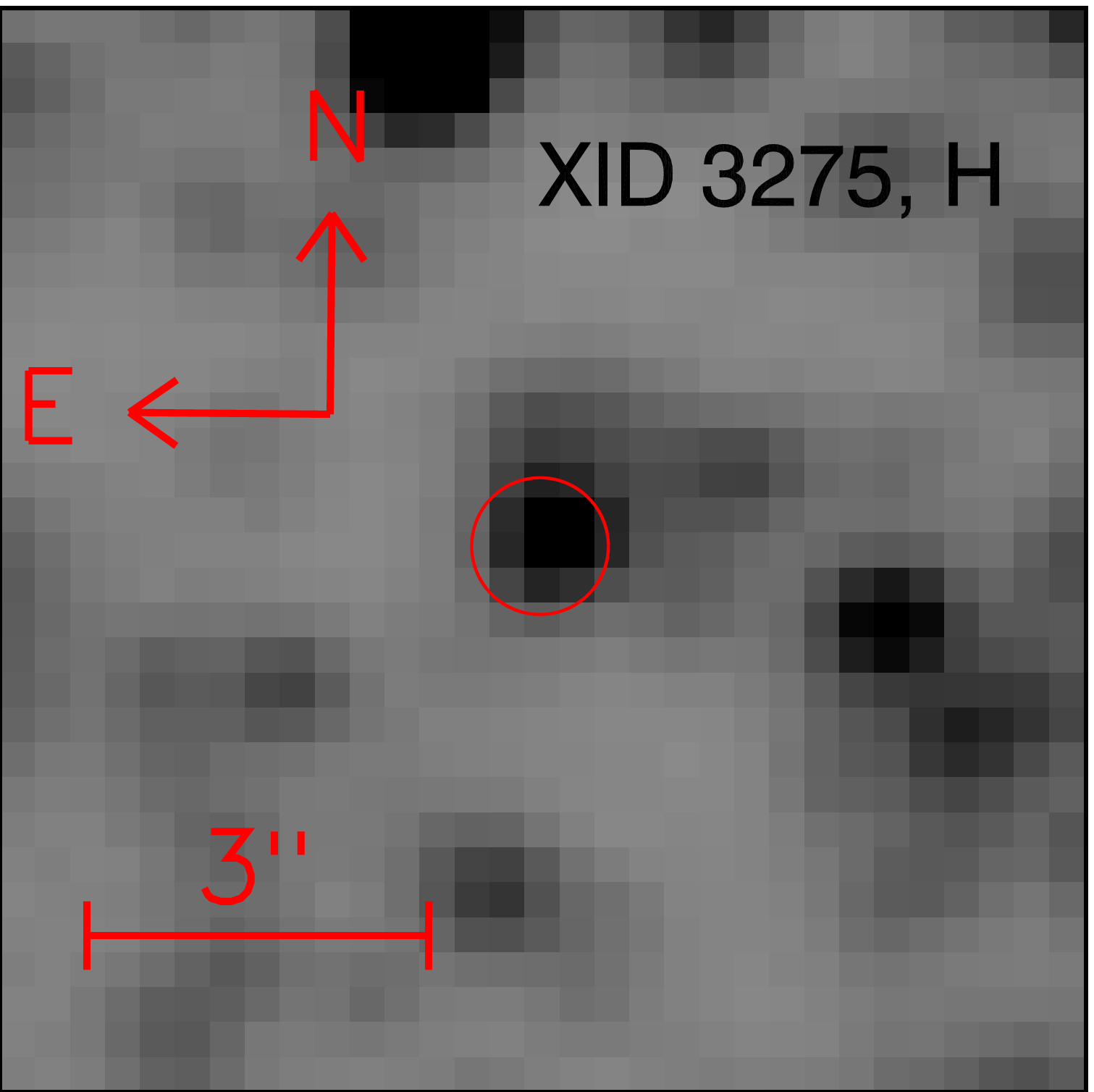}
}
\hspace{0.1cm}
\subfigure[] 
{
    \label{fig:subim_xid3275_Ks}
    \includegraphics[width=3.5cm]{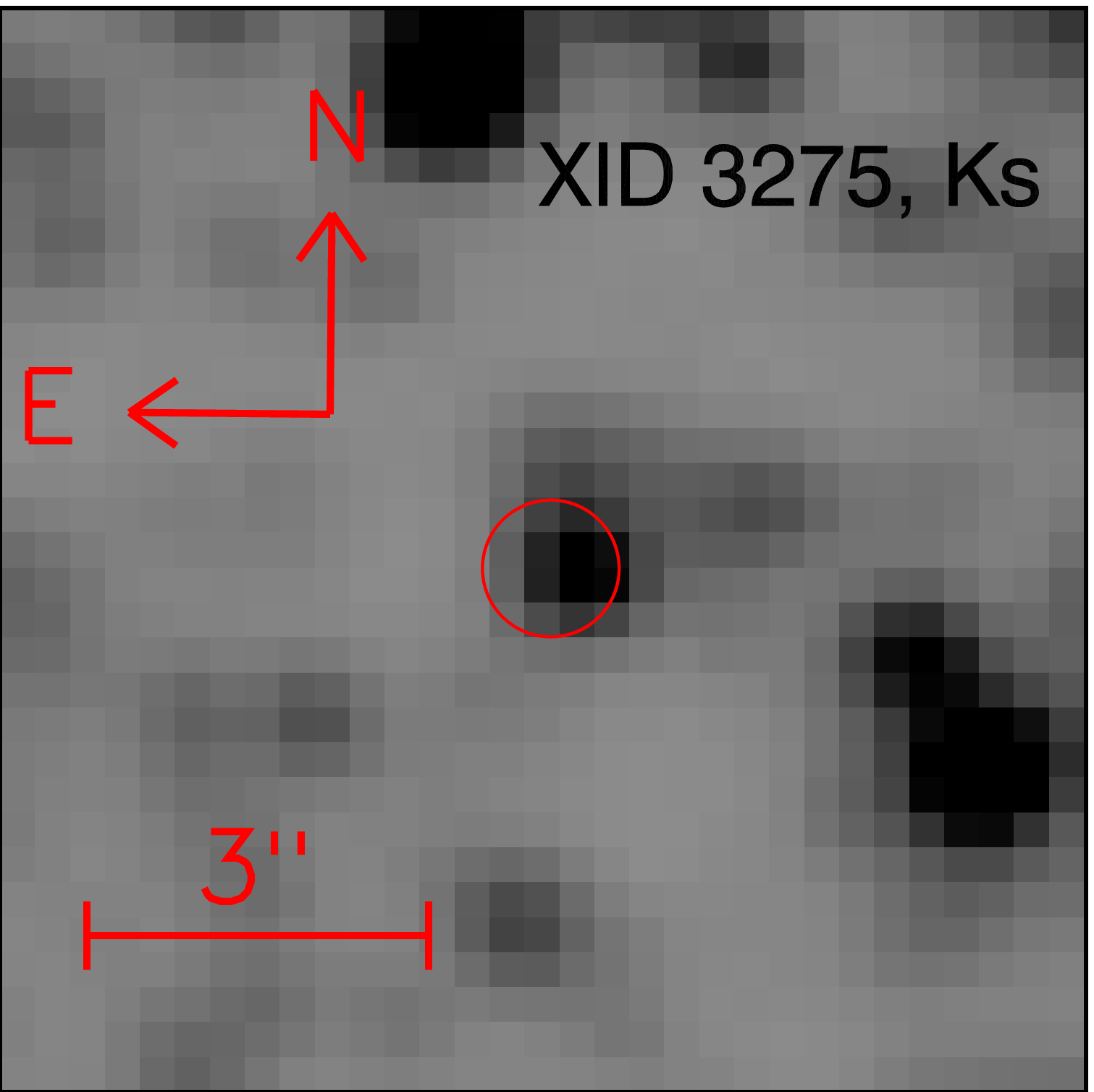}
}

\caption{$10\arcsec~\times~10\arcsec$ finding charts for the counterpart to XID
  3275 from ISPI data. The $95\%$ confidence error circle from
  \cite{muno2009} is shown in red. (a) $J$ band (b) $H$ band (c) $K_{s}$
  band.}
\label{fig:subim_xid3275} 
\end{figure}

\begin{figure}[ht]
\includegraphics[scale=0.75] {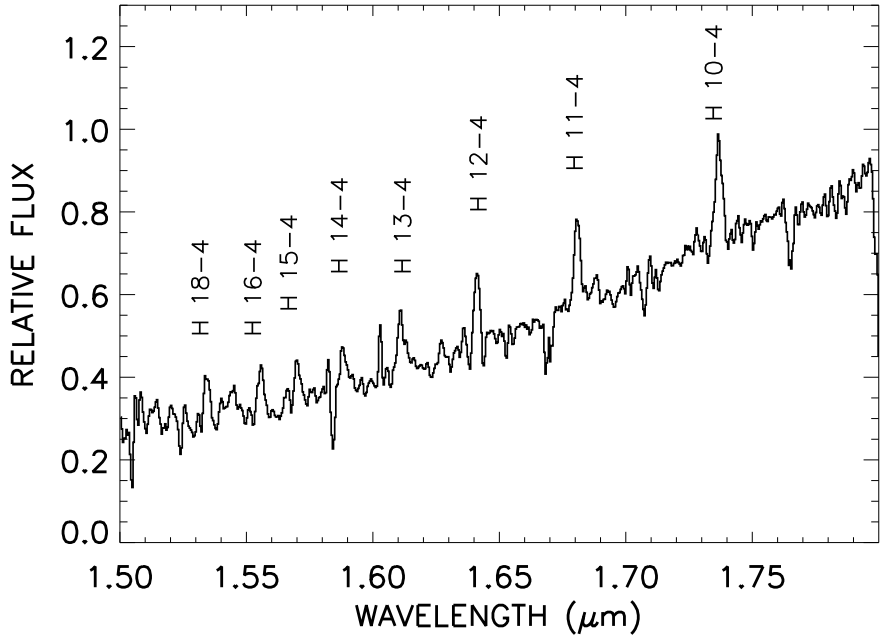}
\caption{LUCIFER1 $H$ band spectrum of the IR counterpart to X-ray
  source XID 3275. The apparent absorption lines are caused by
  incomplete telluric correction.}
\label{fig:xid3275_H}
\end{figure}

\begin{figure}[ht]
\includegraphics[scale=0.75] {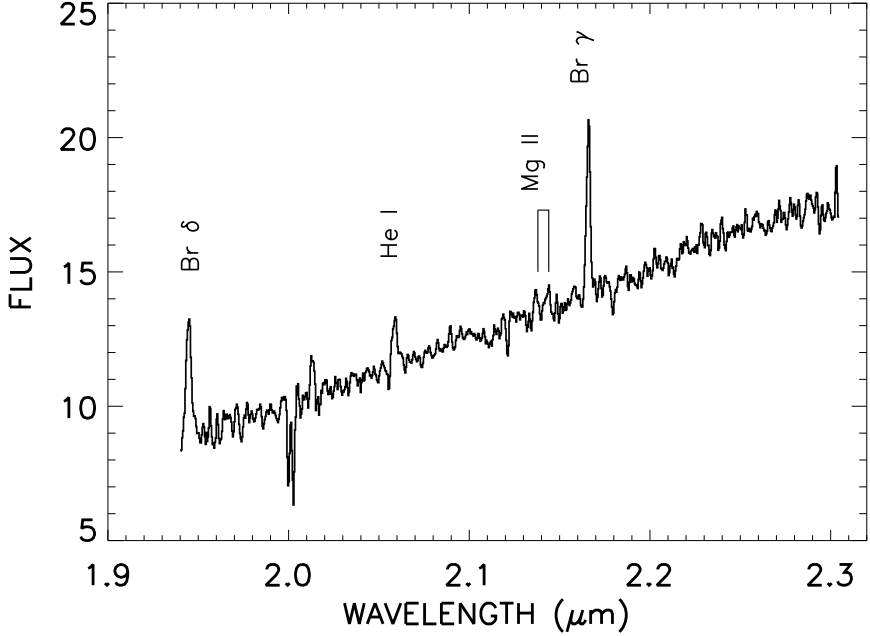}
\caption{LUCIFER1 $K$-band spectrum of the IR
  counterpart to X-ray source XID 3275. The apparent absorption line
  near $\lambda=$2 $\mu$m is caused by incomplete telluric correction.} 
\label{fig:xid3275_K}
\end{figure}

The NIR counterpart to XID 3275 is a $K_{s}=13.08$ mag star with heavy
reddening ($H-K_{s}= 1.67$ mag and $J-K_{s}= 4.61$ mag). Finding
charts produced with ISPI data are shown in
Figure \ref{fig:subim_xid3275}. The spectrum, observed with LUCIFER1,
clearly shows emission from the Brackett series from H 7-4 to H 18-4,
as well as He I 2.0589 $\mu$m (see Figures \ref{fig:xid3275_H} and
\ref{fig:xid3275_K}). There is also evidence for emission from the Mg II
doublet at 2.138 $\mu$m and 2.144 $\mu$m. We quote air wavelengths for
all spectral features. No clear absorption lines are
present in the spectra, and in particular there is no evidence of a CO
feature at 2.295 $\mu$m. At a signal to noise level $S/N=40$ per
resolution element, the absence rules out spectral types later than
F0. Our line identifications and measurements for the NIR spectra are
listed in Table \ref{tab:linesforxid3275}.

\begin{table}[ht]
\caption{Line identifications and parameters for the counterpart
  spectrum of XID 3275.}
\label{tab:linesforxid3275} 
\begin{tabular}{ l l l l l l cr}
\hline
  Line ID   & rest    & observed& W$_{EQ}$        & FWHM        \\
            & wavelength ($\mu$m) & wavelength ($\mu$m)& (\AA) & (\AA)\\
\hline
Br $\gamma$ & 2.16610 & 2.16561$\pm$0.00001 & -11.8 $\pm$ 0.8 & 20  $\pm$ 5 \\
He I        & 2.05870 & 2.05844$\pm$0.00001 & -3.8  $\pm$ 1.0 & 23  $\pm$ 5 \\ 
Br $\delta$ & 1.94510 & 1.94454$\pm$0.00001 & -10.3 $\pm$ 1.5 & 29  $\pm$ 5 \\
Mg II        & 2.13800 & 2.13690$\pm$0.00001  & -2.1   $\pm$  1.0 & 20  $\pm$ 10 \\
Mg II        & 2.14400 & 2.14320$\pm$0.00001  & -1.6    $\pm$ 1.0 &  20 $\pm$ 10 \\
H 10-4      & 1.73670 & 1.73611$\pm$0.00001 & -10.2 $\pm$ 1.5 & 32  $\pm$ 3 \\
H 11-4      & 1.68110 & 1.68053$\pm$0.00001 & -11.4 $\pm$ 1.5 & 32  $\pm$ 5 \\
H 12-4      & 1.64120 & 1.64074$\pm$0.00001 & -8.8  $\pm$ 1.5 & 23  $\pm$ 5 \\
H 13-4      & 1.61140 & 1.61056$\pm$0.00001 & -7.0  $\pm$ 1.5 & 32  $\pm$ 5 \\
H 14-4      & 1.58850 & 1.58822$\pm$0.00001 & -6.8  $\pm$ 1.5 & 23  $\pm$ 5 \\
H 15-4      & 1.57050 & 1.57023$\pm$0.00001 & -8.3  $\pm$ 1.5 & 29  $\pm$ 5 \\
\hline
\end{tabular}
\end{table}

Visually, the spectrum is like that of a typical Be type star. \citet{steele1999}
and \citet{steele2001} present a catalog of $H$ and $K$ band spectra
of 57 optically-typed Be stars between O9-B9. Brackett and He
I emission are identifying characteristics of their classification
for ``Group I'', corresponding to spectral types of O9e-B3e. 
The presence of Mg II emission corroborates this range of spectral types, as it is only
seen in types earlier than B4e. \citep{steele2001}. Therefore we can
constrain the spectral type of this source to O9e-B3e.

Be stars are young, massive O and B stars that are rotating near
their break-up speed, casting off matter into an equatorial
circumstellar disk. The disk material fluoresces as it is irradiated
by UV emission from the stellar photosphere. Although normal Be stars
are not known to be sources of hard X-ray
radiation, Be stars are known for being components in HMXBs and in
$\gamma$ Cas systems, both of which produce hard X-rays
with luminosities detectable at the GC distance. Therefore, we
consider this NIR source to be the likely true physical counterpart to
the X-ray source, XID 3275. 

Be stars are commonly found to be the donor stars in HMXBs. Of the 131
known HMXBs, nearly half are thought to contain Be stars and 28 have
been spectrally confirmed \citep{reig2011}. Be HMXBs are known for
hard X-ray emission described by a power-law. Typically Be XRBs
contain a neutron star in an eccentric orbit with a period of
10-300 days \citep{reig2011}. Accretion episodes occur during the
closest approach of the neutron star and often exhibit X-ray
pulsations due to the neutron star rotation. The high states can be
very luminous, reaching $L_{X}=10^{38}$ ergs s$^{-1}$, while quiescent
luminosities are typically $10^{33}-10^{35}$ ergs s$^{-1}$. In
general, Be HMXBs with fainter and steadier emission are associated
with longer periods and lower orbital eccentricities. 

Given the large range of Brackett $\gamma$ strengths found for
ordinary Be stars ($EW=8~$\AA$~$ in absorption to $EW=-24~$\AA$~$in
emission for the sources in the \cite{steele1999} sample), there is no
clear evidence for or against accretion activity contributing to the
emission features in our spectrum.  The equivalent widths (EW) of the
Brackett $\gamma$ and He I 2.0587 $\mu$m emission lines in the XID 3275
counterpart are -10.3$~$\AA$~$ and -3.8$~$\AA$~$, respectively. These
values are typical for Be stars\citep{clark2000}. 

The emission line FWHMs are consistent with an origin in the
circumstellar material. Using the FWHM$_{Br\gamma}$-$v\sin (i)$
relationship from \citet{steele2001} the FWHM$=$20$~$\AA$~$ translates to
$v \sin (i)=$160 $\pm$ 50 km s$^{-1}$. The Be stars in the Clark sample have a range
of $90$ km s$^{-1}<v \sin (i)< 320$ km s$^{-1}$, which means the line
widths from the spectrum of XID 3275 are typical for Be stars.

\begin{figure}[ht]
\includegraphics[scale=0.75] {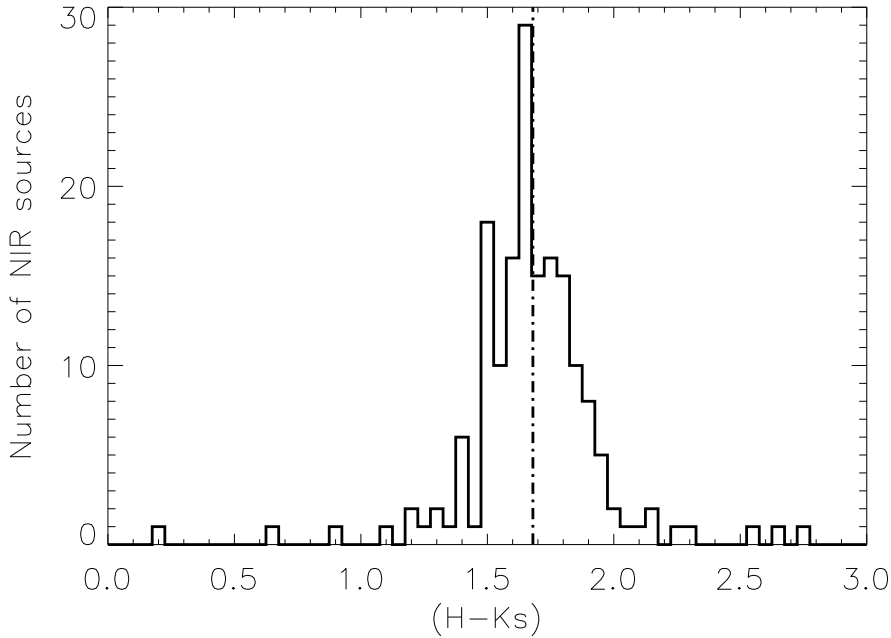}
\caption{The $H-K_{s}$ color of the counterpart to XID 3275 (dashed
  line), compared to the colors of all NIR sources within 20\arcsec$~$of its
  position.}
\label{fig:localcolors3275}
\end{figure}

To determine the luminosity class of this object, we need to constrain
the distance. The color, $(H-K_{s})=1.68$ mag is near the modal value
of the $(H-K_{s})$ distribution of sources within 20\arcsec$~$of this area
(see Figure \ref{fig:localcolors3275}), which implies a distance close
to that of the Galactic Center. O9-B3 stars have intrinsic colors of
$(J-H)_{0}=-0.12$ mag or -0.10 $<~(H-K)_{0}~<$ -0.05 mag, according to
\cite{ducati2001}. However, Be HMXBs are often redder than this due to
circumstellar material; the typical range given in
\citet{reig2011} is -0.2 $<~(J-K)~<$ 0.6 mag, which corresponds to
-0.04 $<~(H-K)~<$ 0.1 mag, approximating the absorption ratios with values
derived from the GC interstellar medium \citep{nishiyama2009}. We adopt
the intrinsic color range -0.04 $<~(H-K)_{0}~<$ 0.1 mag for XID 3275 and a
GC distance value of 8 kpc \citep{reid1993}. Using extinction
coefficients derived for the Galactic Center by \cite{nishiyama2008}
we calculate an extinction range of 2.16 $\le A_{K_{s}} \le$ 2.36 mag. The dereddened
magnitudes are thus $10.56 \le J \le 11.17$ mag, $10.67\le H \le
11.01$ mag and $10.72\le K_{s} \le 10.92$ mag. We calculate an
absolute magnitude of -3.79 $\le$ M$_{K_{s}}\le$ -3.6 mag.
Main sequence O9-B3 stars have absolute magnitudes of
$-0.8<M_{K_{s}}<-3.4$ mag \citep{carroll1996}. For giant stars in the
same range of subtypes, the absolute magnitudes are
$-2.18<M_{K_{s}}<-4.12$ mag. Thus, a range of B0-3e III best
matches the calculated absolute magnitude, but we cannot rule out that our
source is a main sequence star earlier than O9V, or that this distance
and luminosity is not overestimated due to intrinsic reddening in
excess of our assumed maximal value of $(H-K_{s})_{0}=0.1$ mag. However, a
much redder value of $(H-K_{s})_{0}$ would be unusual for a Be HMXB
\citep{reig2011}.

We used XSPEC \citep{arnaud1996} to perform X-ray spectral fitting on
the unbinned X-ray data from \citet{muno2009} using the Cash statistic
(hereafter C-statistic; \citealt{cash1979}), which is more appropriate
for modeling low-count X-ray spectra. We refrain from subtracting the
background, and instead follow \citet{broos2010} to model both source
and background in the observed spectrum. We fit both a \textit{mekal}
thermal bremmstrahlung model \citep{liedahl1995} and an absorbed
power-law model to the X-ray data.

The fits are poorly constrained, but indicate that the source is intrinsically
hard, with $\Gamma=0.4^{+1.3}_{-1.1}$ or kT $>$ 10 keV. The column density from the
model fits is also poorly constrained, but generally consistent with
the values we obtained using NIR colors: $N_{H}=1.1^{+1.1}_{-0.7} \times
10^{23}$ cm$^{-2}$. For the best-fit absorbed power-law model, we
obtained an unabsorbed flux from 0.5-8 keV of $8.0^{+0.7}_{-4.4} \times 10^{-15}$ ergs cm$^{-2}$
s$^{-1}$. At the GC distance of 8 kpc \citep{reid1993} , this corresponds to $6.1^{+0.5}_{-3.4} \times 10^{31}$
ergs s$^{-1}$. 

This measurement would be the second faintest observed X-ray
luminosity for a Be HMXB. \cite{tomsick2011} report that the Be HMXB
IGR J01363+6610 had an upper limit luminosity of L$_{x}< 1.4\times
10^{31}$ ergs s$^{-1}$ during an observation with
\textit{Chandra}. Prior to the \textit{Chandra} observation, the
luminosity of IGR J01363+6610 was measured with \textit{XMM-Newton} to
be L$_{x}=$9 $
\times~ 10^{31}$ ergs s$^{-1}$, also unusually low for a Be
HMXB. At another time, IGR J01363+6610 was observed by INTEGRAL during
an outburst at an X-ray luminosity of $\sim 10^{35}$ ergs s$^{-1}$
\citep{tomsick2011}.

To date, XID 3275 has not been observed during an outburst nor has it
varied outside the \textit{Chandra} noise
error levels despite having been observed 35 times between October
2000 and July 2007 for a total of 11.7 days of on-source time
\citep{muno2009}. We note, however, that there were a number of long
gaps in the coverage: two $\sim 400$ day gaps and two $\sim$ 260 day
gaps during which XID 3275 could have undergone an outburst and
returned to a quiescent state.

The implication could be that XID 3275 is a large separation, low
eccentricity version of the average Be HMXB. Five Be HMXBs are known
to have orbital eccentricities smaller than $e=0.2$ and with $P \ge
30$ days \citep{reig2011}. Due to the lack of close approaches to the
Be star, these systems are observed to have steady luminosities of
$\leq$10$^{33}$ ergs s$^{-1}$. 

Alternatively, XID 3275 may belong to the $\gamma$ Cas subclass of Be
systems. These systems have hard X-ray emission dominated by a $\sim$ 12
keV plasma, instead of the usual hard power-law seen in other Be HMXBs
\citep{lopes2010}. Their luminosity is found in the range of
10$^{32}$-10$^{33}$ ergs s$^{-1}$, which makes them a reasonable
candidate for the X-ray luminosity seen in XID 3275. $\gamma$ Cas
systems also do not display the large amplitude outbursts common to
traditional Be HMXBs. Instead, smaller amplitude variability is seen
on all time scales, from hours to months \citep{lopes2010}. The hourly
flickering has been used by some authors to argue that $\gamma$ Cas
systems host a WD instead of a neutron star, by analogy to the high
temperature thermal plasmas and flickering seen in some CVs
\citep{reig2011}. The largest observed amplitude variability has been
a factor of 3 over time scales of 50-90 days. \citet{muno2009} did not
find evidence for flux variations between the 35 observations of this
source; however we again note the irregularity of the \textit{Chandra}
coverage. The question of whether this system is a classical Be HMXB
or a $\gamma$ Cas may be best determined by modeling the X-ray
spectrum to determine if it is an X-ray power-law (for classical
BeHMXBs) or a thermal plasma (for $\gamma$ Cas systems). However, the
109 X-ray photon spectrum accumulated by \textit{Chandra} over 1 Ms of
exposure time for this
source is insufficient for this distinction. 

\subsection{XID 6592: a Candidate Galactic Center Symbiotic Binary}
\begin{figure}[ht]
\centering
\subfigure[] 
{
    \label{fig:subim_xid6592_J}
    \includegraphics[width=3.5cm]{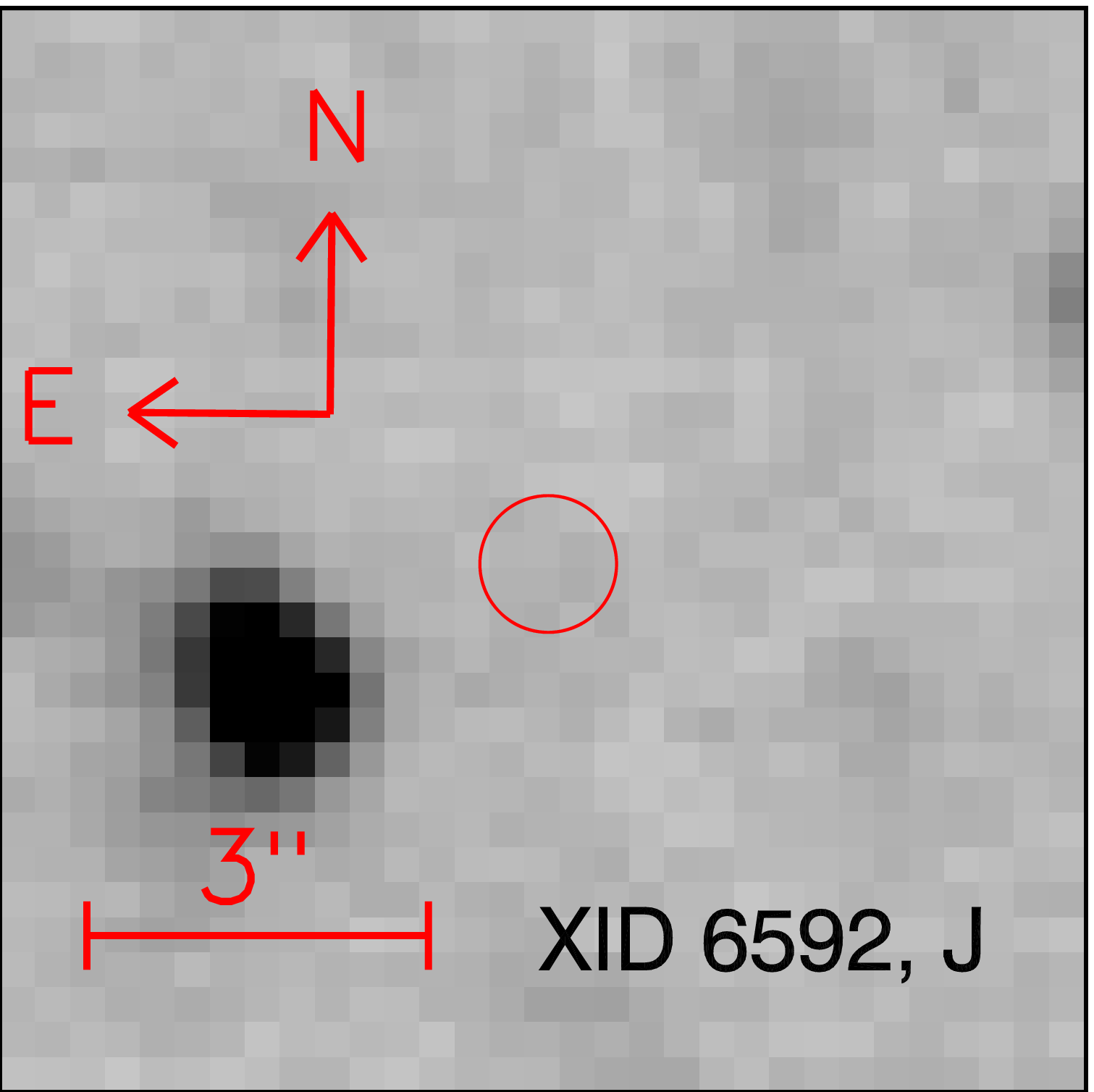}
}
\hspace{0.1cm}
\subfigure[] 
{
    \label{fig:subim_xid6592_H}
    \includegraphics[width=3.5cm]{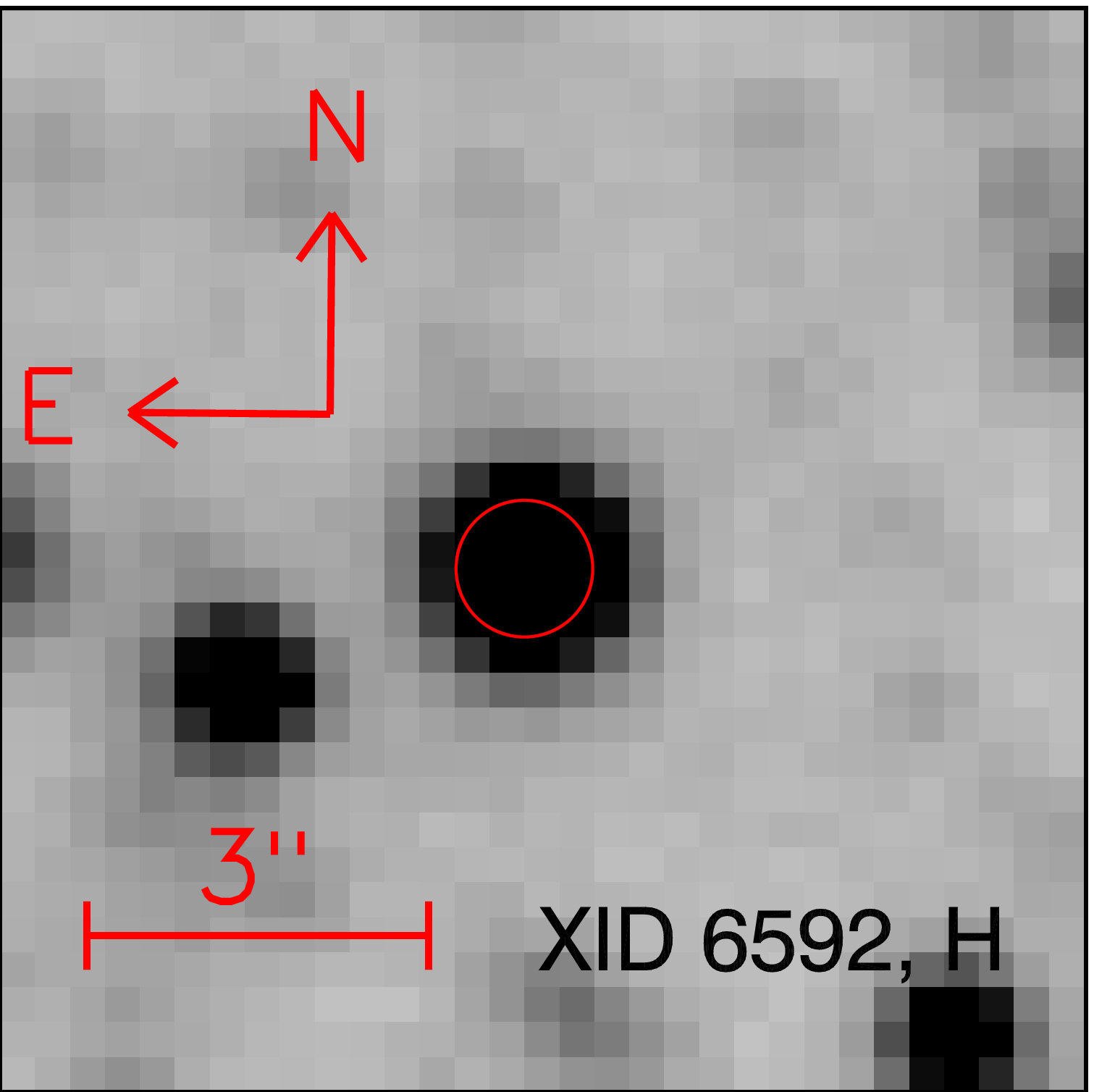}
}
\hspace{0.1cm}
\subfigure[] 
{
    \label{fig:subim_xid6592_Ks}
    \includegraphics[width=3.5cm]{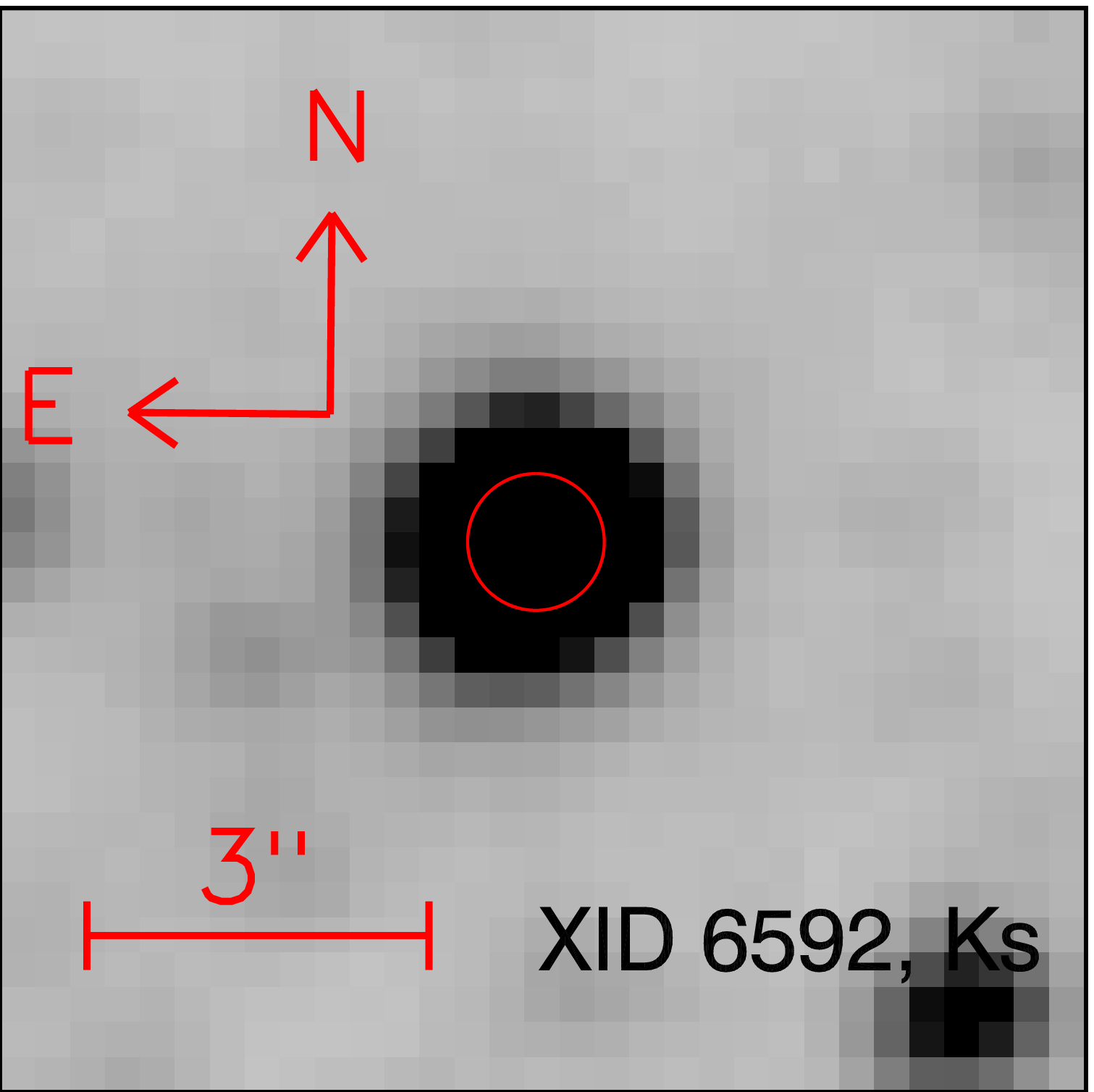}
}

\caption{$10\arcsec~\times~10\arcsec$ finding charts for the counterpart to XID
  6592 produced with ISPI data. The $95\%$ confidence error circle from
  \cite{muno2009} is shown in red. (a) $J$ band (b) $H$ band (c) $K_{s}$
  band.}
\label{fig:subim_xid6592} 
\end{figure}

The counterpart to XID 6592 is a bright $K_{s}=9.88$ mag star with
heavy reddening ($(H-K_{s})=3.15$; undetected in $J$ in our ISPI
data). \cite{matsunaga2009} identify this source as a long period variable based on $H$ and $K_{s}$
photometric monitoring, but they were unable to identify a
period. Finding charts produced with ISPI data are shown in Figure \ref{fig:subim_xid6592}.

\begin{figure}[ht]
\includegraphics[scale=0.75] {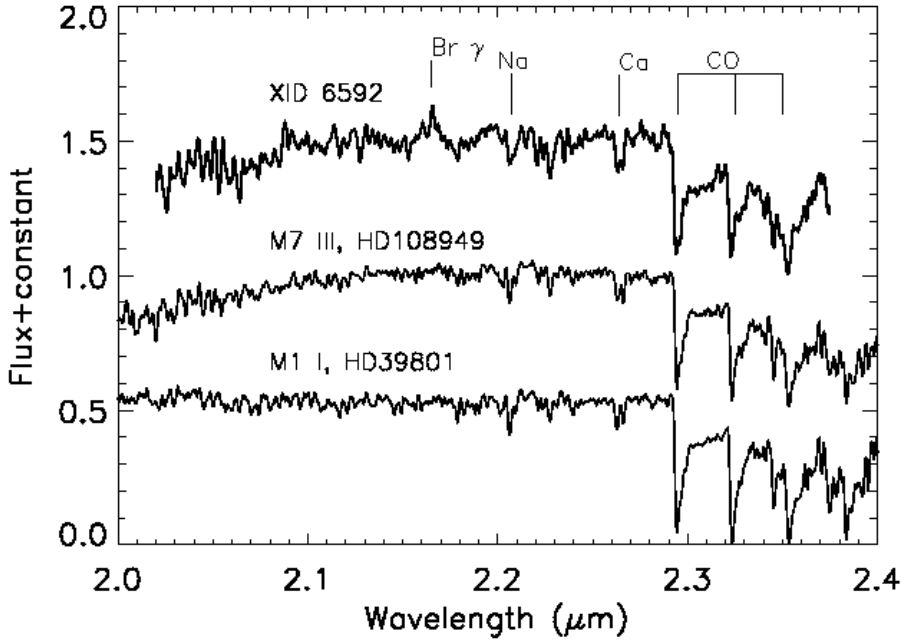}
\caption{OSIRIS $K$-band spectrum of the counterpart to XID 6592. Also shown is
  the best fit M7 III spectrum and the best-fit M 1-2 I spectrum. The
  M I templates all have deeper CO bands than the counterpart to XID 6592,
and therefore we adopt the best-fit M III subtype, M7 III, for this star.} 
  \label{xid6592_spec}
\end{figure}

We identify the counterpart to XID 6952 as a potential X-ray emitting
source on the basis of the large Br $\gamma$ emission line in the $K$ band spectrum, shown in Figure
\ref{xid6592_spec}. However, as we discuss below, there are other
interpretations as well. The $K$ band spectrum shows deep CO bands beyond $\lambda=2.295~\mu$m and deep Na I and Ca I
absorption lines (2.2062/2.2090 $\mu$m and 2.2631/2.2667 $\mu$m,
respectively), as well as a depression in the continuum below
$\lambda=2.09~\mu$m due to steam absorption. 
  Our line measurements from the spectrum are
listed in Table \ref{tab:linesforxid6592}. 

\begin{table}[ht]
\label{tab:linesforxid6592} 
\caption{Line identifications and parameters for the counterpart
  spectrum of XID 6592.}
\begin{tabular}{ l l l l l l cr}
\hline
  Line ID   & rest               & observed& EW        & FWHM        \\
            & wavelength ($\mu$m)& wavelength ($\mu$m)& (\AA) & (\AA)\\
\hline
Br $\gamma$     & 2.16610       & 2.16474$\pm$0.00003 & 1.8$^{+1.0}_{-0.2}$ & 17 $\pm$ 3 \\
Na              & 2.2062/2.2090 & 2.2053/2.2081 $\pm$ 0.0001 &   & - \\ 
Ca              & 2.2631/2.2667 & 2.2619/2.2655 $\pm$ 0.0001 &   & - \\
\hline
\end{tabular}
\end{table}

We derived the spectral type for the counterpart to XID 6592 by calculating
$\chi^{2}$ differences to stellar spectra in the IRTF spectral library
\citep{rayner2009}. First we fit the slope of the continuum between
$2.1~\mu$m and $2.285~\mu$m and removed the slope from both the
counterpart and the IRTF templates. Then we shifted the template
spectra to the velocity of the XID 6592 velocity, using a cross
correlation function. To calculate the $\chi^{2}$ value, we assumed
the template spectrum had negligible noise and used the standard
deviation between the counterpart spectrum and a smoothed version of
itself as the noise value.  We calculated the $\chi^{2}$ value in the
region between $\lambda=2.24~\mu$m and $\lambda=2.35~\mu$m, which
contains the segment of the CO band acquired by OSIRIS, as well as the
Ca feature at 2.285 $\mu$m (Table \ref{tab:chifit6592}).

The best matches in terms of $\chi^{2}$ value for XID 6592 were a
semi-regular variable M7 III star (HD108849)
and the M supergiants. However, in the supergiants, the CO band minima overshoot the levels in XID 6592 by 15 $\%$ or
more.  XID 6592 would also be underluminous for a supergiant at the GC
distance (see below), which leads us to conclude that the M7 III giant
is the most likely spectral type.

\begin{figure}[ht]
\centering
\subfigure[] 
{
    \label{fig:color:a}
    \includegraphics[width=7cm]{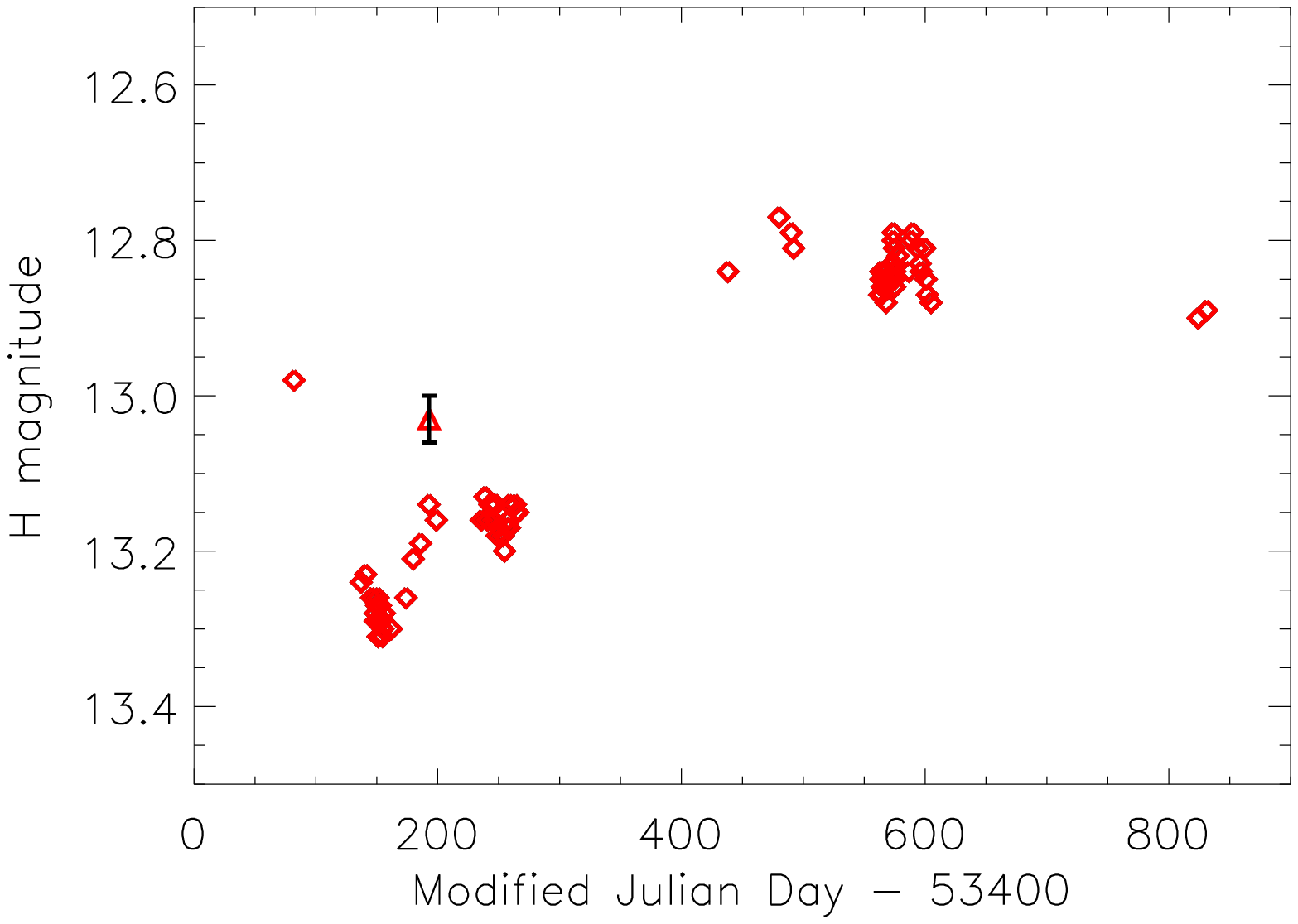}
}
\hspace{0.1cm}
\subfigure[] 
{
    \label{fig:colorimage:b}
    \includegraphics[width=7cm]{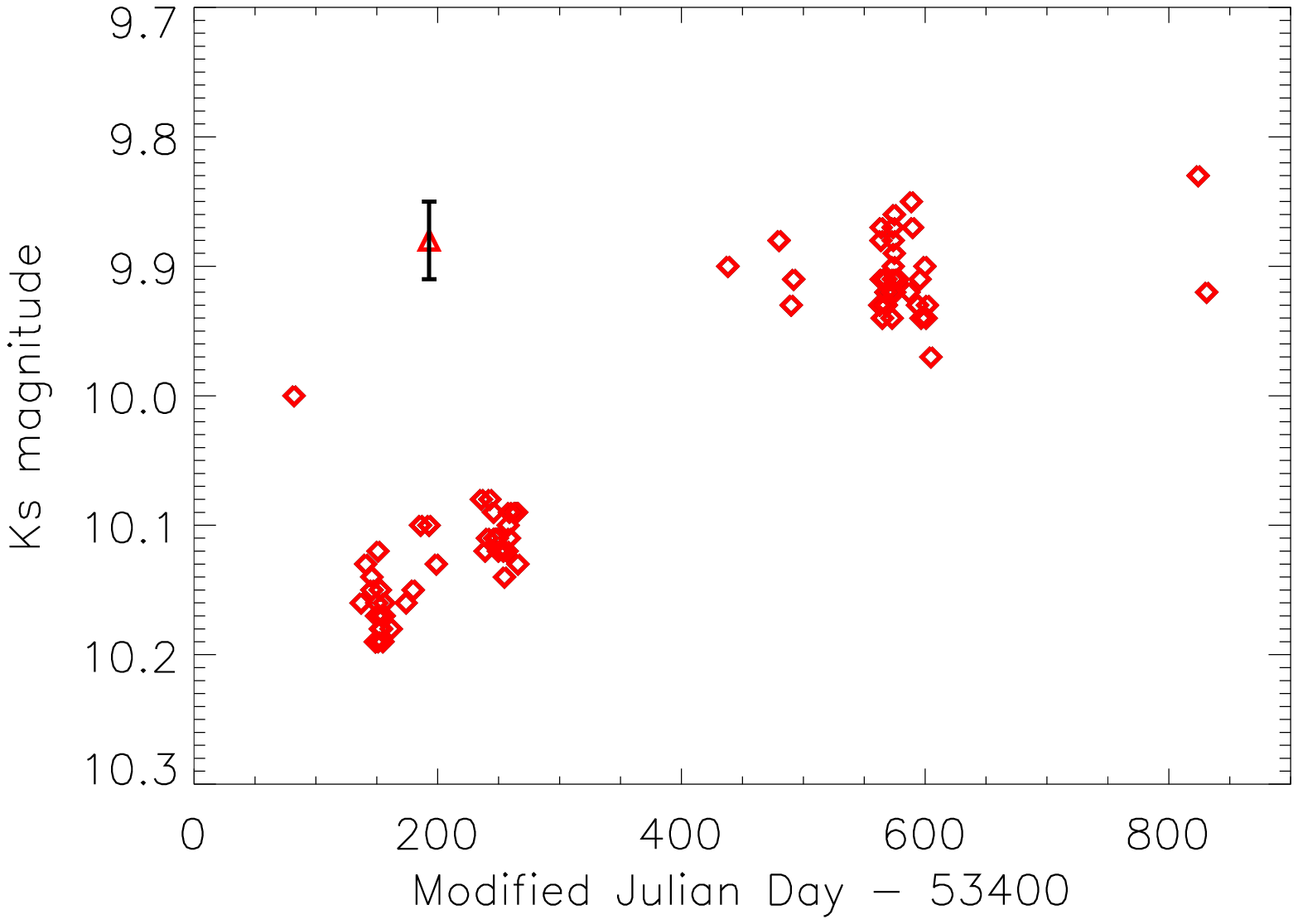}
}
\caption{$H$ and $K_{s}$ band light curve of the NIR counterpart of
  XID 6592 observed with SIRIUS \citep{matsunaga2009} plotted with
  diamonds. The ISPI measurement is overplotted using
  a triangle symbol.}
\label{fig:lightcurvescolorimage} 
\end{figure}

Our ISPI data falls within the time span of the photometric monitoring
reported in \citet{matsunaga2009}. In Figure
\ref{fig:lightcurvescolorimage} we show their $H$ and $K_{s}$
lightcurves of the XID 6592 counterpart, with the addition of our ISPI
photometry. The ISPI $K_{s}$ band and $H$ band seem to be
outliers from the SIRIUS $H$ and $K_{s}$ light curve; both our
measurements are significantly brighter and may indicate a flaring
episode in the data. We would not insist on this interpretation, since
the source is very bright and our measurement could be affected by
pixel non-linearity. Detailed comparison of our ISPI photometry and
the SIRIUS photometry from \citet{matsunaga2009} should be performed to verify a possible
flaring episode.

\begin{table}[ht]
\begin{center}
\label{tab:chifit6592} 
\caption{$\chi^{2}$ fit for XID 6592 counterpart.}
\begin{tabular}{ l l l l}
\hline
  Sp.Type   & $\chi^{2}$  (D.O.F.$=$149) & reduced $\chi^{2}$ & template ID \\
\hline
K2 III      & 7913.5   & 53.1 & HD132935       \\
K6 III      & 5229.5   & 35.1 & HD3346         \\
M1 III      & 3579.2   & 24.0 & HD204724       \\
M4 III      & 2764.4   & 18.6 & HD214665       \\
M6 III      & 1119.1   & 7.5  & HD196610       \\ 
M7 III      &  721.3   & 4.8  & HD108849       \\ 
M5e-M9e III & 1129.5   & 7.6  & HD14386 (Mira) \\
M9 III      & 1069.7   & 7.2  & IRAS15060+0947 \\
            &          &      &                \\
M0.5Ib      & 1242.2   & 8.3  & HD236697       \\
M1-2 Ia-Iab & 695.6    & 4.7  & HD39801        \\
M3-4 Iab    & 718.3    & 4.8  & HD14469        \\
M5 Ib-II    & 2061.6   & 13.8 & HD156014       \\
\hline
\end{tabular}
\end{center}
\end{table}

The intrinsic color for the template M7 III is $(H-K)_{0}=$0.414 mag
\citep{rayner2009}. We use the XID 6592 counterpart mean color from
the observations of \citet{matsunaga2009} ($(H-K_{s})=2.89$ mag), and
an extinction scaling of $A_{V}:A_{H}:A_{K_{s}}=16.15:1.74:1$
\citep{nishiyama2008,nishiyama2009}, to calculate A$_{K_{s}}=3.3$ mag
and A$_{V}=53.2$ mag for the extinction to XID 6592. 

\begin{figure}[ht]
\includegraphics[scale=0.75] {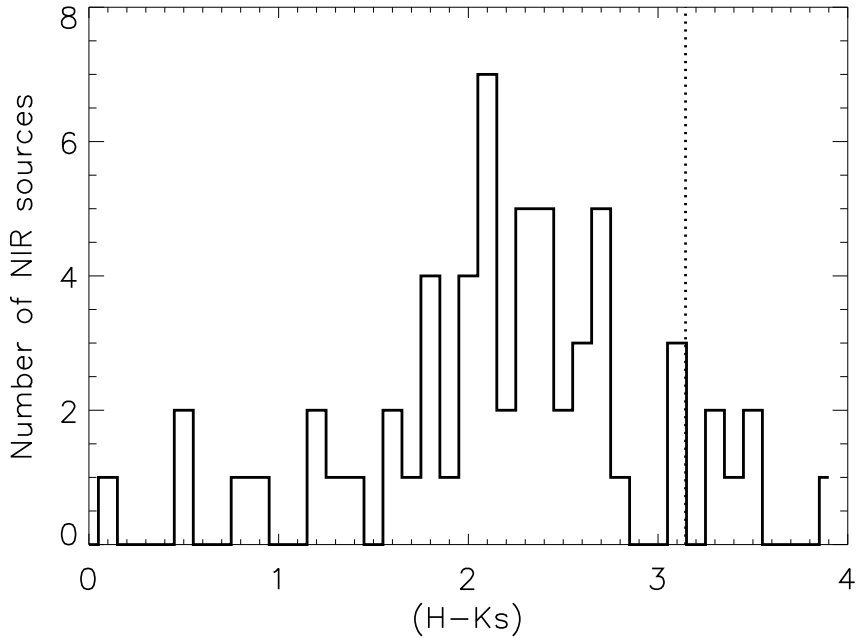}
\caption{The $H-K_{s}$ color of the counterpart to XID 6592, compared
  to the colors of all NIR sources within 20\arcsec$~$of the target
  position.}
\label{fig:localcolors6592}
\end{figure}

Normal late type stars should dominate the Galactic Center
population and have colors of $(H-K_{s}) < 0.2$ mag. Even
adjusting for the intrinsic color difference, this source's reddening
is larger by at least 0.7 mag than the modal value for the local
extinction of this region (Figure \ref{fig:localcolors6592}), so it
may either lie beyond the GC distance, or be subject to intrinsic
extinction in addition to what is present in the best fit template
(meaning that $(H-K)_{0}~>~$0.414 mag).  We adopt the GC distance of 8
kpc \citep{reid1993} but note that it may represent the minimum distance to this
object. For the GC distance we calculate an absolute magnitude
$M_{K_{s}}=-7.91$ mag, given a spectral type of M7 III. This value
falls within the range of GC LPVs studied by
\citet{blum2003}, which range from $-9.78 \le M_{K} \le -7.76$ mag.

Using the ratio of column density to the visual extinction
($\frac{N_{H}}{A_{V}}=1.79 \times 10^{21}$ cm s$^{-1}$ mag$^{-1}$
; \citealt{predehl1995}), we estimate that XID 6592 lies behind
$N_{H}=9.6~\times~10^{22}$ cm$^{-2}$. We also fit the \textit{Chandra}
data for this source from \citet{muno2009} using XSPEC with an
absorbed powerlaw \citep{arnaud1996}. We found that XID 6592 is absorbed by
$N_{H}=2.55_{-1.13}^{+0.88} \times 10^{23}$ cm$^{2}$. This is 2.5 times larger
than the estimate derived from NIR colors, which may indicate
intrinsic absorption of the X-ray source. The power-law
slope was found to be $\Gamma=1.5_{-1.2}^{+0.9}$ and the best-fit
parameters yield an unabsorbed X-ray flux (0.2-8 keV) of $F_{X}=4.2^{+0.2}_{-0.2}
\times 10^{-13}$ ergs cm$^{-2}$ s$^{-1}$. For the GC distance of 8 kpc
\citep{reid1993}, this
corresponds to $L_{X}=3.2^{+0.1}_{-0.1} \times 10^{33}$ ergs s$^{-1}$.

The counterpart spectrum shown in Figure \ref{xid6592_spec} shows a
Brackett $\gamma$ emission line at 2.1645 $\mu$m, with
an equivalent width of -1.8 to -2.6$~$\AA$~$, depending on the
placement of the continuum. This corresponds to a blue shift of
$-80\pm 20$ km s$^{-1}$ with respect to the prominent
Na and Ca absorption lines.

Isolated Mira variables are not known to
be strong X-ray sources. However, they often show emission lines from
hydrogen during some phases of their pulsations. This is thought to be
due to shocks forming in the expanding and contracting envelope of the
star. \citet{hinkle1984} found that Mira emission lines are
blueshifted an average of 13.2 km/s with respect to the absorption
lines of the star. The maximum emission line velocity found for Miras
in their study was $-18$ km s$^{-1}$.  We could not find many
measurements of the EW for
Brackett emission from isolated Miras in the literature. The
\citet{wallace1997} spectral atlas includes 9 spectra of o Ceti (Mira)
taken at different phases across the pulsation period. We measured the
Brackett $\gamma$ emission line EWs of the atlas
spectra of o Ceti from different pulsational phases and found
it to vary between 0 and −1.2±0.2$~$/AA, smaller than the EW we measure
for XID 6592's spectrum.

\citet{chakrabarty1998} measured the blueshifted emission in the
symbiotic binary GX 1+4 and found velocities up to $-250\pm70$
km s$^{-1}$ for He I 1.083 $\mu$m. GX 1+4 is known to contain a
pulsar. \citet{chakrabarty1998} considered whether the blueshifted
emission line could originate from the ionization of the wind of the M
giant, or from a wind driven directly from the pulsar or accretion
disk. Since M giant wind speeds are expected to have a range of 10$-$30
km s$^{-1}$ \citep{dupree1986}, they conclude in favor of an outflow
driven by the pulsar or accretion disk.

Our value of $-80\pm 20$ km s$^{-1}$ is inconsistent with the
known velocity values of self-shocked emission from Miras. Therefore
we believe that there is a binary companion to the red giant in the
XID 6592 system that is either ionizing the red giant's wind or
directly driving an outflow. This companion, if compact, could also be
responsible for the X-ray emission from the system. The high velocity
seen in the Br $\gamma$ line is more consistent with outflow driven
from an accretion disk around the binary companion than by ionization
of wind from the red giant. 

Twenty percent of the objects in the \citet{belczynski2000} symbiotic binary
catalog have irregular variables or Mira variables as secondary
stars. Their mass loss at this evolutionary stage makes them effective
mass donors to any compact objects in $<100$ day orbits
\citep{murset1999}.The remainder of symbiotic secondaries consist of
normal first ascent M type giants.

Most symbiotic stars discovered so far contain white dwarf compact
objects. The majority of these sources show X-rays characteristic of
0.4$-$1 keV plasmas, as a result of quasi-steady nuclear burning of
accreted material on the WD surface \citep{murset1997}. The exceptions
are NS symbiotic binaries also known as symbiotic X-ray binaries and the newly emerging class of hard
X-ray emitting white dwarf symbiotic binaries. 

The WD or NS identification is often made by X-ray spectral fits which
simultaneously solve for accretion rate (by measuring the internal
absorption of the X-rays) and total X-ray luminosity. The amount of
energy liberated by accretion onto a compact object is inversely proportional to
its surface radius. Thus, once the accretion rate and resulting
luminosity are known, the radius of the compact object can be
constrained \citep{luna2007}. The radius of a white dwarf has a
negative dependence on the mass \citep{provencal1998}, allowing X-ray
spectral fitting to solve for the WD mass \citep{luna2007}.

There are several symbiotic binaries now known to host neutron star
primaries, including GX 1+4, 4U 1700+24, 4U 1954+31, IGR J16194-2810,
IGR 16393-4643 and IGR 16358-4726 \citep{luna2007,nespoli2010}. These
objects display X-ray luminosities between 10$^{32}$-10$^{34}$ ergs
s$^{-1}$ \citep{masetti2002,nespoli2010}. GX 1+4 has been observed in
outburst at 10$^{37}$ ergs s$^{-1}$ \citep{staubert1995}.

\citet{masetti2006} took optical spectra of the NS symbiotic binaries 4U
1954+319 and 4U 1700+24 during periods of relatively low X-ray
luminosity of $\sim$10$^{32}$ ergs s$^{-1}$. They found the optical spectra
for both sources were completely devoid of emission lines, with the
continuum dominated by the secondary M III stars. The lack of Balmer
lines in the optical make it unclear whether these sources would have
detectable Brackett emission in the NIR. \citet{nespoli2010} took
$K$-band spectra of IGR 16393-4643 and IGR 16358-4726 and found tenuous
detections of Brackett $\gamma$ and He I 2.0581 $\mu$m in each. The spectrum
of GX 1+4 taken in the $K$ band by \citet{chakrabarty1998} has strong
emission at Brackett $\gamma$ and no other emission features.

There are at least 13 known hard spectrum symbiotic binaries thought to host
white dwarfs, including RT Cru, TCrB, CH Cyg, CD-573057, SS73 17 and 8
others recently detected by the X-ray telescope on the \textit{Swift
  Gamma Ray Observatory}
\citep{luna2007,smith2008,luna2010}. \citet{luna2007} model the X-ray
spectra of RT Cru with the purpose of finding the source of the hard
X-ray emission. They exclude magnetic channeling of the accretion
flow, as seen in hard spectrum WD polars, on the basis of the lack of
pulsations caused by the WDs rotation. Instead they find that the WD
mass is unusually high, $\sim 1.3$M$_{\sun}$, which causes higher energy
collisions at the surface of the WD than for lower mass WDs. White
dwarf masses are usually 0.55$-$1 M$_{\sun}$ in symbiotic binaries
\citep{sion1992}.

Similarly high white dwarf masses are calculated for the other systems
in this class \citep{luna2008,smith2008,luna2010}. \citet{eze2011}
suggest that this source class could be a major channel for Type Ia
Supernovae. SN Ia's are usually thought to result from the collapse of
Chandrasekhar mass ($\sim$1.4M$_{\sun}$) CO white dwarfs. Various
accreting binaries with white dwarfs have been proposed as the
progenitors, but most of these systems are either too rare for the
observed SN Ia rate or unlikely to accrete enough mass to reach the
Chandrasekhar mass limit \citep{vankerkwijk2010}. Hard spectrum WD
symbiotic binaries may signify systems that have nearly succeeded in reaching
the Chandrasekhar mass and thus may prove to be a major channel for
SN Ia \citep{eze2011}.

The X-ray luminosity range found in hard X-ray WD symbiotic binaries is
between 10$^{32}$-10$^{34}$ ergs s$^{-1}$ \citep{luna2007,smith2008}.
To date we know of no published NIR spectra of any of these sources,
but the optical spectra are known to have Balmer line emission in at
least three cases \citep{cielslinski1994,mukai2003}. This gives us
reason to expect Brackett line emission in the NIR from hard spectrum
WD symbiotic binaries.

Thus, we have two strong candidates for the nature of this system: NS
symbiotic X-ray binaries and hard-spectrum WD symbiotic binaries. Each of these candidates
have the luminosity range and hard spectral characteristics of XID
6592. The nature of the compact object could in principle be
constrained with observations of NS bursts or pulsations, or by X-ray
spectral modeling; however, the present tally of 150 photon counts,
acquired in $\sim$ 60 ks of exposure time, is insufficient to draw
firm conclusions. Additional X-ray observations may be warranted. 

\subsection{XID 947: a Candidate Galactic Center O star}

\begin{figure}[ht]
\centering
\subfigure[] 
{
    \label{fig:subim_xid947_J}
    \includegraphics[width=3.5cm]{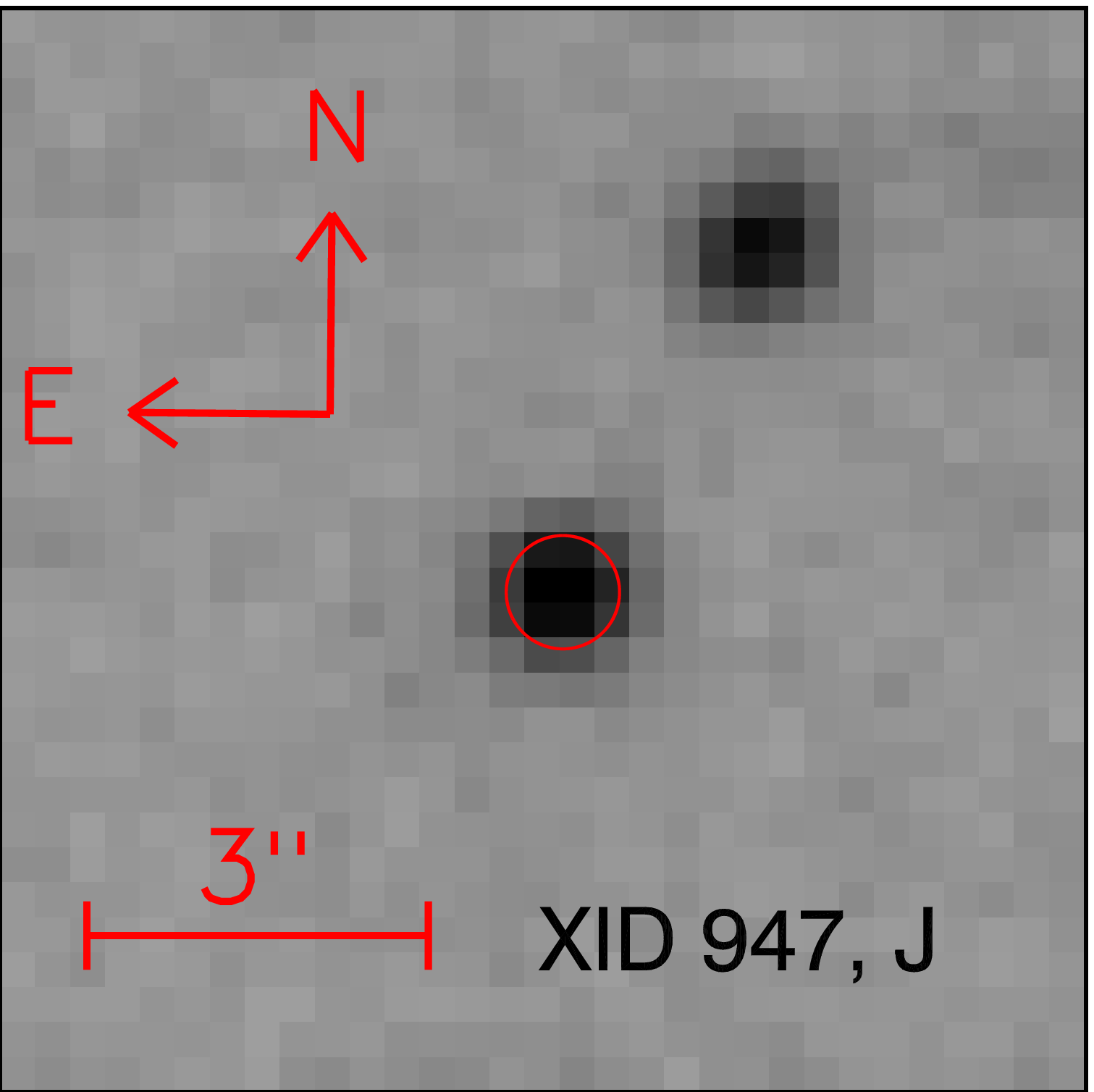}
}
\hspace{0.1cm}
\subfigure[] 
{
    \label{fig:subim_xid947_H}
    \includegraphics[width=3.5cm]{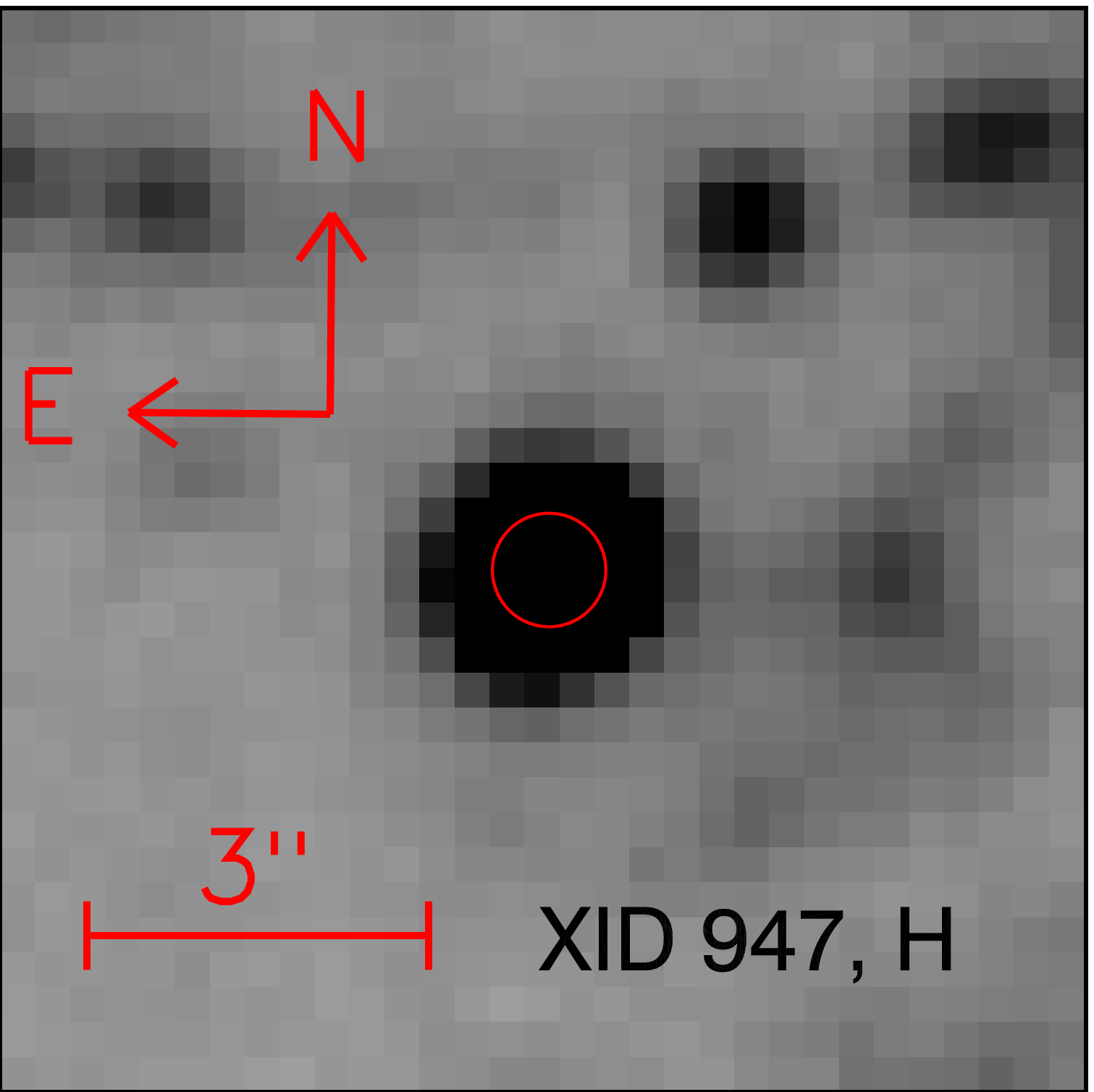}
}
\hspace{0.1cm}
\subfigure[] 
{
    \label{fig:subim_xid947_Ks}
    \includegraphics[width=3.5cm]{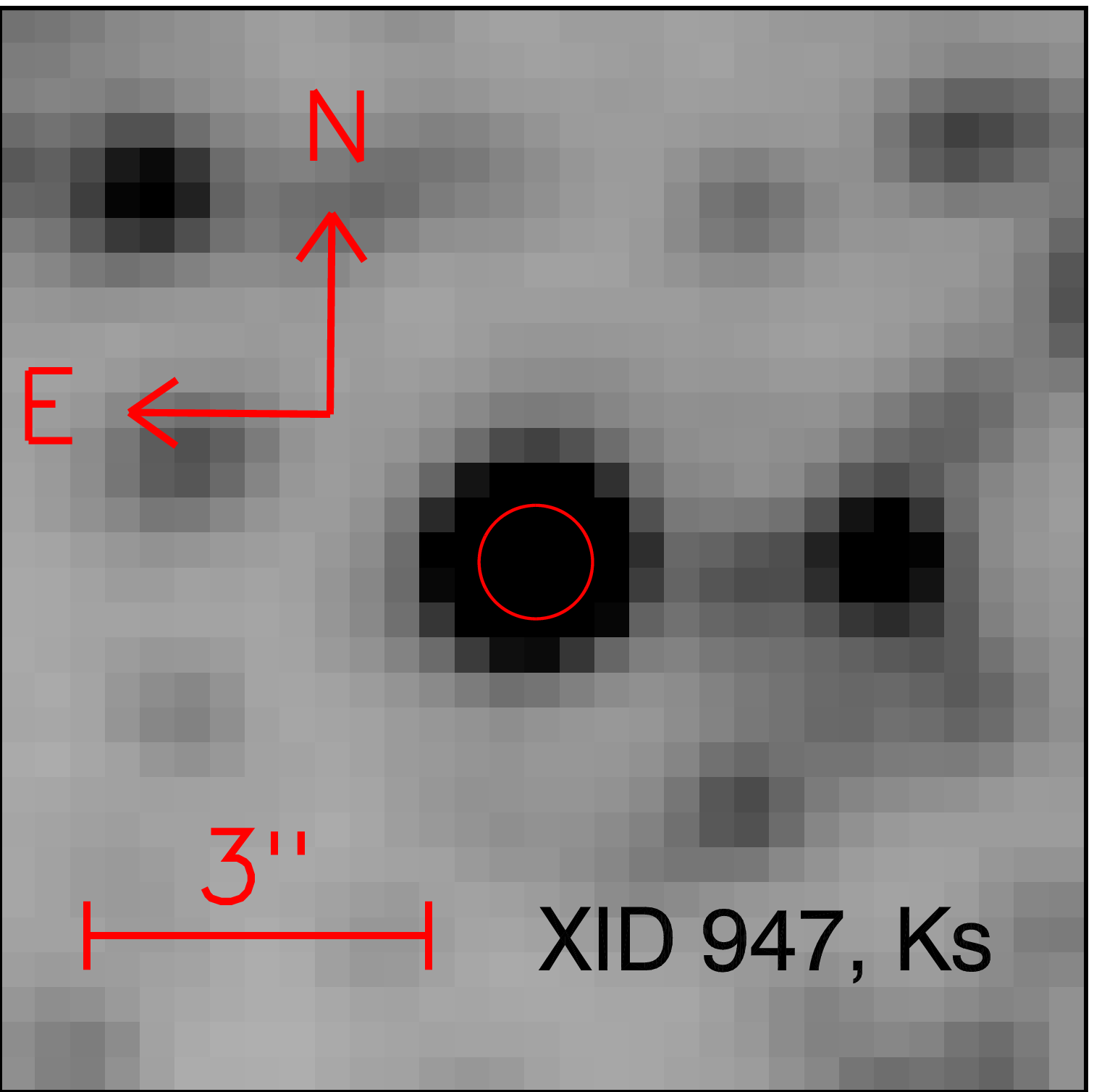}
}
\caption{$10\arcsec~\times~10\arcsec$ finding charts for the counterpart to XID
  947 from ISPI data. The $95\%$ confidence error circle from
  \cite{muno2009} is shown in red. (a) $J$ band (b) $H$ band (c) $K_{s}$
  band.}
\label{fig:subim_xid947} 
\end{figure}

\begin{figure}[ht]
\includegraphics[scale=0.75] {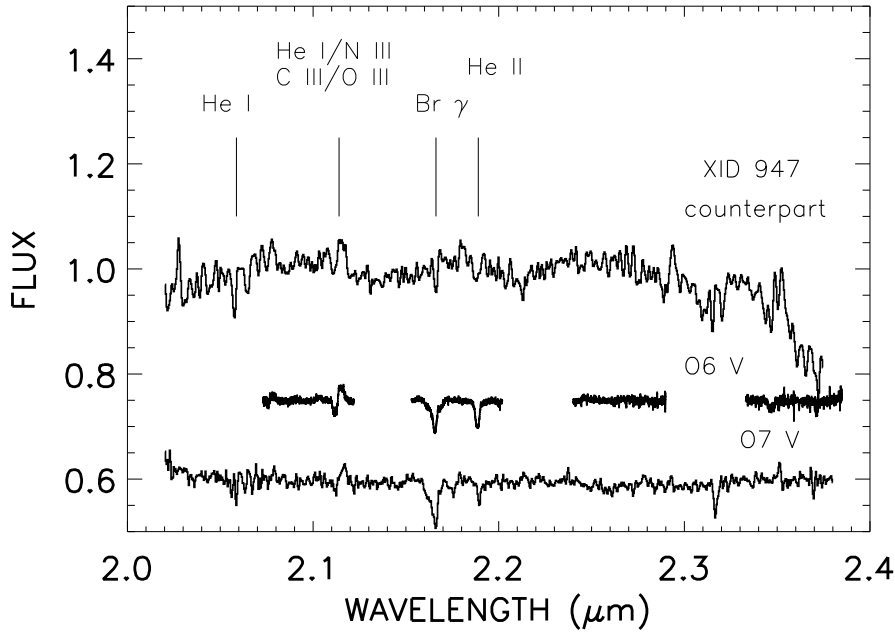}
\caption{OSIRIS $K$-band spectrum of the counterpart to XID 947. Also
  plotted are an O6 V spectrum from \citet{hanson2005} and an O7 V from
 \citet{wallace1997}.}
\label{fig:xid947spec}
\end{figure}

The counterpart to XID 947 is a $J=16.23$ mag, $H=13.29$ mag,
$K_{s}=11.61$ mag star shown in Figure \ref{fig:subim_xid947}. We
observed the $K$-band spectrum with OSIRIS. We identify it as a
probable true counterpart because of the lack of apparent CO features
at $\lambda>2.295$ $\mu$m and because of the marginal detections of an He
I/N III P-Cygni feature at 2.113 $\mu$m and weak Brackett $\gamma$
absorption commonly found in mid O-type spectra
\citep{hanson2005}. This spectrum is difficult to type beyond the lack
of CO absorption because the telluric transmission was poorly
corrected by the two observations of the A1 V star taken before and
after this target (see Figure \ref{fig:xid947spec}). This may be due to
the fact that the slit position angle of the telluric reference
spectra was taken near the parallactic angle, and $\sim$ 100$^{\circ}$
away from the position angle used for the observation of XID 947. The
spectrum of this source may therefore be affected by
wavelength-dependent slit losses, exacerbated by the high airmass at the time of
the observations [$sec(z)=2.1$].

\begin{figure}[ht]
\includegraphics[scale=0.75] {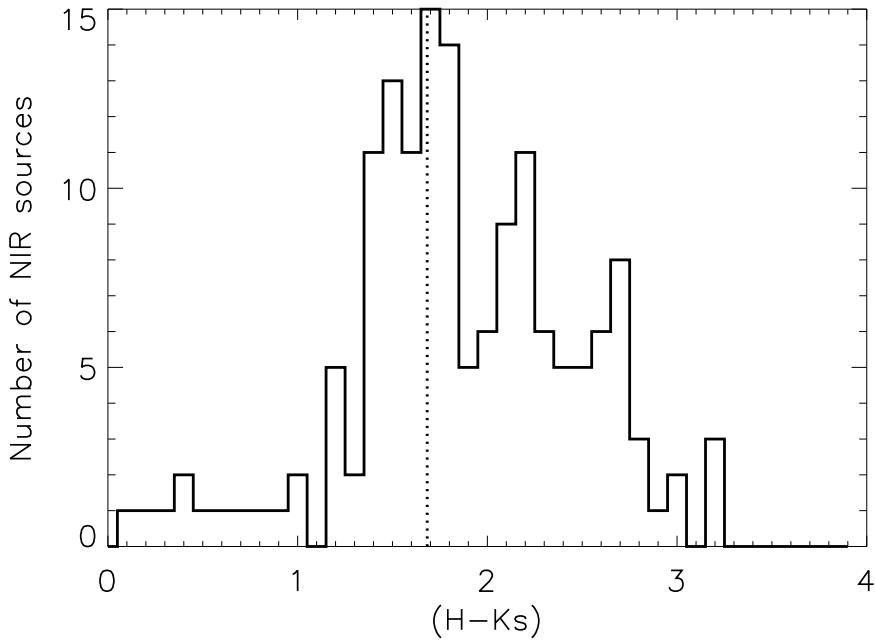}
\caption{The $H-K_{s}$ color of the counterpart to XID 947, compared
  to the colors of all NIR sources within 20\arcsec$~$of the target
  position.}
\label{fig:localcolors947}
\end{figure}

The $(H-K_{s})$ color of the XID 947 counterpart is at the modal value
of the color distribution of other NIR sources within 20\arcsec (see
Figure \ref{fig:localcolors947}). Therefore we adopt the Galactic
Center distance of 8.0 kpc for this source \citep{reid1993}.

We proceed by adopting the intrinsic colors of O stars from
\citet{martins2006}: $(H-K)=-0.10$ mag and $(J-H)=-0.11$ mag.  We adopt
$\Delta(H-K_{s})=1.78$ mag for the reddening and calculate
$A_{J}=7.26$ mag, $A_{H}=4.16$ mag, $A_{K_{s}}=2.40$ mag and
$A_{V}=38.8$ mag using extinction scalings from \citet{nishiyama2008}.
The extinction-to-column-depth conversion ratio
$\frac{N_{H}}{A_{V}}=1.79\times10^{21}$ cm$^{-2}$ gives us a column
depth value of $N_{H}=6.9\times10^{22}$ cm$^{-2}$ toward XID 947 \citep{predehl1995}.
We calculate absolute magnitudes of $M_{J}=-5.55$ mag,
$M_{H}=-5.39$ mag, $M_{K_{s}}=-5.31$ mag. This magnitude is compatible
with O3-9.5 I supergiants, O6 III giants, and O3 V dwarfs
\citep{martins2006}.

We also modeled the \textit{Chandra} data with XSPEC
\citep{arnaud1996} using both an absorbed thermal bremsstrahlung model
and an absorbed power law model. The bremsstrahlung model yields an
unrealistically high temperature of $kT=39$ keV. The absorbed
power-law model has a best-fit parameters of
$\Gamma=-0.1^{+1.0}_{-0.9}$ and $N_{H}=4.04^{+4.9}_{-3.6} \times
10^{22}$ cm$^{-2}$. For these parameters, the unabsorbed X-ray flux is
$3.4^{+0.5}_{-1.3} \times 10^{-15}$ ergs cm$^{-2}$ s$^{-1}$, which corresponds to
$2.6^{+0.4}_{-1.0} \times 10^{31}$ ergs s$^{-1}$ at the GC distance. 

This is on the low end for the
known O supergiants that are detected in the GC
\citep{mauerhan2010}. A single O V or O III star could in principle
generate this luminosity in X-ray photons, from internal shocks within
its wind. The typical scaling of the X-ray flux to the bolometric
luminosity is $\frac {L_{X}}{L_{bol}}=10^{-7}$ for isolated O stars of
all luminosity classes \citep{oskinova2005}. For a $K_{s}$-band bolometric
correction of -4.4 \citep{mauerhan2010b}, $log(L_{bol})=-5.5$. This
yields a ratio $\frac {L_{X}}{L_{bol}}$ of 10$^{-7.5}$, which is lower than the canonical value for isolated O stars
\citep{oskinova2005}. However, our XSPEC spectral model calculated the
total flux of only a power-law component, extrapolating the flux to
soft X-ray energies that were absorbed in our spectrum. If there was
an additional soft thermal X-ray component, the resultant flux would
also be absorbed but it would not be taken into account by our XSPEC
luminosity calculation. Most of the sources used to derive the
canonical $\frac {L_{X}}{L_{bol}}=10^{-7}$ value had multiple soft, thermal X-ray components \citep{oskinova2005}. Therefore, XID 947 may
have a typical X-ray luminosity for an O star, when all spectral
components are taken into account.

We conclude that XID 947 is likely to be an O-type star with uncertain
evolutionary class that may be isolated, in a colliding-wind
binary, or quietly accreting onto a compact object. A
higher S/N ratio spectrum is needed to make any significant
conclusions about this source.

\section{Conclusion}
Our spectroscopic observations found three new counterparts to the
population of X-ray sources near the Galactic Center, increasing the
number identified in the inner $17'\times17'$ by 20$\%$. These included
a Be HMXB or $\gamma$ Cas system, a symbiotic X-ray binary, and an O-type star with an undetermined
luminosity class. 

Despite observing 17 of our ``high probability'' candidate
counterparts, the Be HMXB  XID 3275 was the only one to exhibit
emission lines in the NIR spectra. The GC O-type star and symbiotic XRB
show emission lines but were not part of the ``high probability'' set due to lower X-ray
flux and a lack of a $J$-band detection, respectively. 

The remaining 16 ``high probability'' targets were all late-type
giants without emission lines. We expected to discover 6$\pm$2 true
counterparts to X-ray sources based
on the 38$\pm$9$\%$ probability found in our simulations \citep{dewitt2010}. 
We believe that our previous statistical analysis was robust, and that this
apparent discrepancy may be due to observing candidate counterparts
during phases of low accretion activity and/or that the nature of
these sources is such that accretion
signatures are not generally visible in their spectra, as in the case
for some symbiotic X-ray binaries (e.g. \citealt{masetti2006,nespoli2010}).
We will address these possibilities and their implications for the GC
X-ray source population in a
future paper.

For the candidate Be HMXB/$\gamma$ Cas system (XID 3275), we inferred a spectral type of
B0-3 III. We do not have enough well-sampled X-ray data to determine if there is X-ray variability of
the expected type for a classical Be HMXB or a $\gamma$ Cas system;
the \citet{muno2009} data are consistent with constant X-ray flux over a long time
baseline. We also note that this source would be atypically X-ray
faint for a Be HMXB, although such a faint X-ray luminosity is not
unprecedented. Further regular X-ray monitoring of this source to
search for X-ray variability is required to determine its nature.

The O type star, XID 947 could be a colliding wind binary or a
high mass X-ray binary or an isolated star with unusually energetic
X-ray emission. This source adds to the
total of 32 similar hard X-ray spectrum massive young stars (WR/O types) toward the Galactic
Center.

Finally, we discovered a hard X-ray emitting likely symbiotic star
with an M7 III semi-regular variable secondary (XID 6592). It has a Brackett $\gamma$ line
blueshifted by 80 km s$^{-1}$ with respect to the stellar center of velocity
for the M giant, which we interpret as a feature from an accretion
driven wind. Symbiotic stars are usually known for having low energy
X-ray spectra with kT$<$1 keV. The hard X-ray spectrum for this source
could be explained by a neutron star companion, which tends to produce
a high-energy power-law tail, or this source could be a member of the
newly emerging class of nearly Chandrasekhar mass white dwarf
symbiotic binaries. These stars are of interest as possible
progenitors to supernovae Type Ias.

\pagebreak
The authors thank Richard Pogge, Matthias Dietrich and Chris Howk for
performing the observations of our targets with LBT LUCIFER1. 

CD, RMB, and SSE gratefully acknowledge the support of NSF grant 
AST-0807867.

OSIRIS (Ohio State Infrared
Imager/Spectrometer) is a collaborative project between The
Ohio State University and Cerro Tololo Inter-American Observatory
(CTIO) and was developed through NSF grants AST 90-
16112 and AST 92-18449.

CTIO is part of the National Optical Astronomy Observatory (NOAO), based in La Serena, Chile.
NOAO is operated by the Association of Universities for Research
in Astronomy (AURA), Inc., under cooperative agreement
with the National Science Foundation. 

This work has been
done with observations from the Southern Astrophysics Research
(SOAR) telescope, a collaboration among the Ministério
da Ciência e Tecnologia/Brazil, NOAO, The University of
North Carolina at Chapel Hill, and Michigan State University.

FEB acknowledges the support of CONICYT, Chile, under grants FONDECYT
1101024 and FONDAP (CATA) 15010003. 

\appendix
\section{Appendix: Table of All Sources Spectroscopically Observed}

In Table \ref{tab:nonemitters} we include the coordinates and ISPI photometry of all NIR counterparts that we targeted as part of our survey. Except where otherwise noted, these sources had no emission
lines corroborating X-ray activity. Every deeply reddened non-emission
line NIR source [$(H-K_{s}) > 1.0]$ had spectral features indicating a
late type. Foreground sources [$(H-K_{s})<1.0$] had a mix of spectral
types, but no emission lines indicating high activity. 

\begin{table}[!ht]
\begin{threeparttable}
\tiny
\setlength{\tabcolsep}{0.05in}
\caption{Near-infrared counterparts to X-ray sources included in our spectroscopic
  survey.  Bold face source names indicate that the source was one of our ``high probability'' targets.}
\label{tab:nonemitters}
{\begin{tabular}{ l l l l l l l l l l l l l cr}
\hline
  \textit{Chandra} X-ray   & X-ray         & RA$_{IR}$   & DEC$_{IR}$ & $J$        & $H$      & $K_{s}$ & Observatory & Comments\\
    Source (CXOUGC)      & Catalog No.  & (J2000.0)  & (J2000.0)   & (mag)& (mag)& (mag) & &\\
\hline
 J174459.0-285123 & XID 6010 &17:44:59.05 &-28:51:23.6 &  15.65$\pm$ 0.05   &14.19$\pm$ 0.04 &12.94$\pm$ 0.05& OSIRIS/SOAR&\\
 J174459.4-285140 & XID 6019 &17:44:59.41 &-28:51:40.7 &  13.44$\pm$ 0.02   &11.90$\pm$ 0.03 &10.72$\pm$ 0.03& OSIRIS/SOAR&\\
\bf{J174459.5-285108} & XID 6023 &17:44:59.53 &-28:51:08.0 &   13.88$\pm$ 0.03  &12.32$\pm$ 0.03 &11.26$\pm$ 0.04& OSIRIS/SOAR&\\
 J174500.0-285132 & XID 6034 &17:45:00.08 &-28:51:32.9 &  13.88$\pm$ 0.03   &12.93$\pm$ 0.03 &12.23$\pm$ 0.04& OSIRIS/SOAR&\\
 J174500.8-285121 & XID 6055 &17:45:00.82 &-28:51:21.9 &  -                           &15.28$\pm$ 0.07 &12.83$\pm$ 0.05& OSIRIS/SOAR&\\
 J174507.4-290456   & XID 57 &17:45:07.41 &-29:04:57.3  &  -                            &14.47$\pm$ 0.04 &12.52$\pm$0.04& OSIRIS/SOAR&\\
\bf{J174510.1-290515} & XID 6262 &17:45:10.18 &-29:05:15.2 &   17.74$\pm$ 0.17  &14.15$\pm$ 0.04 &12.23$\pm$ 0.04& OSIRIS/SOAR&\\
 J174527.8-290109  & XID 861 &17:45:27.86 &-29:01:09.4 &   14.90$\pm$ 0.04   &11.02$\pm$ 0.03 &  8.95$\pm$ 0.03& OSIRIS/SOAR&\\
 J174528.0-290023  & XID 885 &17:45:28.02 &-29:00:23.0 &   14.46$\pm$ 0.03   &12.59$\pm$ 0.03 &11.48$\pm$ 0.04& OSIRIS/SOAR&\\
 J174528.5-285959  & XID 930 &17:45:28.55 &-28:59:59.6 &   15.86$\pm$ 0.05   &11.98$\pm$ 0.03 &  9.90$\pm$ 0.03& OSIRIS/SOAR&\\
 J174528.7-290942 & XID 6592 &17:45:28.79 &-29:09:42.8 &  -                           &13.03$\pm$ 0.03 & 9.88$\pm$  0.03& OSIRIS/SOAR& symbiotic XRB; described above\\
 J174528.8-285726  & XID 947 &17:45:28.88 &-28:57:26.4 &   16.23$\pm$ 0.06   &13.29$\pm$ 0.03 &11.61$\pm$ 0.04& OSIRIS/SOAR& O-type star; described above\\
 \bf{J174529.0-290406}  & XID 978 &17:45:29.03 &-29:04:06.5 &    14.88$\pm$ 0.04  &10.45$\pm$ 0.03 &  8.52$\pm$ 0.03& OSIRIS/SOAR&\\
 J174529.6-290227 & XID 1030 &17:45:29.68 &-29:02:27.2 &  17.12$\pm$ 0.10   &12.85$\pm$ 0.03 &10.61$\pm$ 0.03& OSIRIS/SOAR&\\
\bf{J174530.3-290341} & XID 1100 &17:45:30.27 &-29:03:41.8 &   17.49$\pm$ 0.14  &13.59$\pm$ 0.04 &11.46$\pm$ 0.04& OSIRIS/SOAR&\\
 J174531.6-290048 & XID 1229 &17:45:31.65 &-29:00:48.3 &  17.63$\pm$ 0.14   &13.77$\pm$ 0.04 &11.79$\pm$ 0.04& OSIRIS/SOAR&\\
 J174532.4-290126 & XID 1330 &17:45:32.47 &-29:01:26.0 &  15.64$\pm$ 0.05   &13.57$\pm$ 0.04 &12.53$\pm$ 0.04& OSIRIS/SOAR&\\
 J174533.5-285540 & XID 1444 &17:45:33.53 &-28:55:39.9 &  -                           &14.39$\pm$ 0.04 &11.94$\pm$ 0.04& OSIRIS/SOAR&\\
 J174533.5-290759 & XID 6687& 17:45:33.44 &-29:07:59.7 &  -                             &13.77$\pm$ 0.04 &10.91$\pm$ 0.03& OSIRIS/SOAR&\\
\bf{J174533.7-285728} & XID 1470 &17:45:33.76 &-28:57:28.7 &   18.45$\pm$ 0.23  &14.55$\pm$ 0.05 &12.46$\pm$ 0.04& OSIRIS/SOAR&\\
 J174536.4-290227 & XID 1783 &17:45:36.43 &-29:02:28.0 &   17.29$\pm$ 0.12  &12.23$\pm$ 0.03 &  9.46$\pm$ 0.03& OSIRIS/SOAR&\\
 J174537.9-290134 & XID 1944 &17:45:37.99 &-29:01:34.5 &  10.78$\pm$ 0.02   & 9.54$\pm$  0.03 & 8.76$\pm$  0.03& OSIRIS/SOAR& O4-6 I; \citep{mauerhan2010b} \\
\bf{J174540.1-290055} & XID 2212 &17:45:40.16 &-29:00:55.5 &   16.78$\pm$ 0.08  &12.09$\pm$ 0.03 &  9.52$\pm$ 0.03& OSIRIS/SOAR&\\
\bf{J174543.4-285841} & XID 2553 &17:45:43.51 &-28:58:41.1 &   17.90$\pm$ 0.18  &15.39$\pm$ 0.07 &13.77$\pm$ 0.07& LBT/LUCIFER&\\
 J174544.4-285829 & XID 2642 &17:45:44.47 &-28:58:29.9 &   -                          &13.26$\pm$ 0.03 &10.54$\pm$ 0.03& LBT/LUCIFER&\\
\bf{J174544.6-285806} & XID 2663 &17:45:44.70 &-28:58:06.6 &    14.85$\pm$ 0.04 &12.96$\pm$ 0.03 &11.94$\pm$ 0.04& LBT/LUCIFER&\\
\bf{J174546.9-285903}& XID 2870 &17:45:46.88 &-28:59:03.2 &     16.27$\pm$ 0.06 &13.23$\pm$ 0.03 &11.73$\pm$ 0.04& LBT/LUCIFER&\\
 J174547.2-285816 & XID 2890 &17:45:47.27 &-28:58:16.2 &    17.81$\pm$ 0.17 &14.96$\pm$ 0.06 &13.02$\pm$ 0.05& LBT/LUCIFER&\\
 J174547.7-285719 & XID 2931 &17:45:47.75 &-28:57:19.8 &    18.09$\pm$ 0.22 &15.01$\pm$ 0.06 &13.38$\pm$ 0.06& LBT/LUCIFER&\\
 J174549.6-285704 & XID 3077 &17:45:49.61 &-28:57:04.1 &    18.41$\pm$ 0.23 &15.08$\pm$ 0.06 &13.35$\pm$ 0.06& LBT/LUCIFER&\\
 J174550.5-290201 & XID 3145 &17:45:50.53 &-29:02:01.6 &  15.41$\pm$ 0.04   &11.83$\pm$ 0.03 &  9.87$\pm$ 0.03& OSIRIS/SOAR&\\
 J174551.2-285838 & XID 3178 &17:45:51.21 &-28:58:39.2 &    14.27$\pm$ 0.03 &13.53$\pm$ 0.04 &13.06$\pm$ 0.05& LBT/LUCIFER&\\
 J174552.0-285507 & XID 3218 &17:45:52.07 &-28:55:06.6 &    16.82$\pm$ 0.08 &12.90$\pm$ 0.03 &10.84$\pm$ 0.03& LBT/LUCIFER&\\
\bf{J174552.9-285537} & XID 3275 &17:45:52.97 &-28:55:37.0 &   17.69$\pm$ 0.16  &14.75$\pm$ 0.05 &13.08$\pm$ 0.05& LBT/LUCIFER& B0-3e III;described above\\
 J174554.0-285432 & XID 3334 &17:45:54.11 &-28:54:33.1 &    18.78$\pm$ 0.25 &15.45$\pm$ 0.07 &13.82$\pm$ 0.08& LBT/LUCIFER&\\
\bf{J174554.4-285455} & XID 3360 &17:45:54.47 &-28:54:55.7 &    18.26$\pm$ 0.21 &14.90$\pm$ 0.06 &12.96$\pm$ 0.05& LBT/LUCIFER&\\
 J174554.8-285650 & XID 3390 &17:45:54.82 &-28:56:50.5 &    12.15$\pm$ 0.02 &11.98$\pm$ 0.03 &11.87$\pm$ 0.04& LBT/LUCIFER&\\
 J174556.0-285056 & XID 7204 &17:45:56.02 &-28:50:56.4 &   16.60$\pm$ 0.07  &13.52$\pm$ 0.04 &12.12$\pm$ 0.04& OSIRIS/SOAR&\\
 J174556.3-285054 & XID 7214 &17:45:56.35 &-28:50:54.0 &  15.08$\pm$ 0.04   &13.76$\pm$ 0.04 &13.31$\pm$ 0.06& OSIRIS/SOAR&\\
\bf{J174557.3-285353} & XID 3496 &17:45:57.34 &-28:53:53.9 &   16.06$\pm$ 0.06  &12.61$\pm$ 0.03 &10.79$\pm$ 0.03& LBT/LUCIFER&\\
 J174558.1-285303 & XID 7266 &17:45:58.19 &-28:53:03.8 &  -                           &13.79$\pm$ 0.04 &11.54$\pm$ 0.03& LBT/LUCIFER&\\
\bf{J174558.6-285511} & XID 3547 &17:45:58.73 &-28:55:11.2 &   18.32$\pm$ 0.22  &15.43$\pm$ 0.07 &13.90$\pm$ 0.08& LBT/LUCIFER&\\
 J174559.7-290112 & XID 3593 &17:45:59.79 &-29:01:12.8 &   15.66$\pm$ 0.04  &11.99$\pm$ 0.03 &10.10$\pm$ 0.03& OSIRIS/SOAR&\\
\bf{J174604.8-285439} & XID 7440 &17:46:04.74 &-28:54:38.7 &   17.28$\pm$ 0.12  &13.40$\pm$ 0.03 &11.47$\pm$ 0.04& LBT/LUCIFER&\\
\bf{J174605.5-285319} & XID 3768 &17:46:05.59 &-28:53:20.2 &   15.30$\pm$ 0.04  &11.47$\pm$ 0.03 &  9.48$\pm$ 0.03& OSIRIS/SOAR&\\
\bf{J174606.3-285313} & XID 7474 &17:46:06.39 &-28:53:14.6 &   16.10$\pm$ 0.06  &12.95$\pm$ 0.03 &11.37$\pm$ 0.04& LBT/LUCIFER&\\
\bf{J174607.6-285351} & XID 7497 &17:46:07.72 &-28:53:52.3 &   16.75$\pm$ 0.08  &13.44$\pm$ 0.04 &11.75$\pm$ 0.04& LBT/LUCIFER &\\
\hline
\end{tabular}}
\end{threeparttable}
\end{table}
\pagebreak

\bibliography{bibliography}
\end{document}